\documentclass[twocolumn]{aastex631}

\usepackage{graphicx}	
\usepackage{amsmath}	
\usepackage{multirow}
\shorttitle{SN~2018gj}
\shortauthors{Teja et al.}

\begin{document}

\title{SN~2018gj: A Short-plateau Type II Supernova with Persistent Blue-shifted $\rm H\alpha$ Emission}

\correspondingauthor{Rishabh Singh Teja}
\email{rishabh.teja@iiap.res.in, rsteja001@gmail.com}

\author[0000-0002-0525-0872]{Rishabh Singh Teja}
\affiliation{Indian Institute of Astrophysics, II Block, Koramangala, Bengaluru-560034, Karnataka, India}
\affiliation{Pondicherry University, R.V. Nagar, Kalapet, Pondicherry-605014, UT of Puducherry, India}

\author[0000-0003-2091-622X]{Avinash Singh}
\affiliation{Hiroshima Astrophysical Science Center, Hiroshima University, Higashi-Hiroshima, Hiroshima 739-8526, Japan}

\author[0000-0002-6688-0800]{D.K. Sahu}
\affiliation{Indian Institute of Astrophysics, II Block, Koramangala, Bengaluru-560034, Karnataka, India}

\author[0000-0003-3533-7183]{G.C. Anupama}
\affiliation{Indian Institute of Astrophysics, II Block, Koramangala, Bengaluru-560034, Karnataka, India}

\author[0000-0001-7225-2475]{Brajesh Kumar}
\affiliation{Aryabhatta Research Institute of Observational Sciences, Manora Peak, Nainital-263001, Uttarakhand, India}

\author{Tatsuya Nakaoka}
\affiliation{Hiroshima Astrophysical Science Center, Hiroshima University, Higashi-Hiroshima, Hiroshima 739-8526, Japan}

\author[0000-0001-6099-9539]{Koji S Kawabata}
\affiliation{Hiroshima Astrophysical Science Center, Hiroshima University, Higashi-Hiroshima, Hiroshima 739-8526, Japan}

\author[0000-0001-9456-3709]{Masayuki Yamanaka}
\affiliation{Kagoshima University, Amanogawa Galaxy Astronomy Research Center, 1-21-35, Korimoto, Kagoshima 890-0065, Kyushu-Okinawa, Japan}

\author[0000-0003-1423-5516]{Takey Ali}
\affiliation{National Research Institute of Astronomy and Geophysics (NRIAG), Helwan 11421, Cairo, Egypt}

\author[0000-0002-4540-4928]{Miho Kawabata}
\affiliation{Department of Astronomy, Kyoto University, Kitashirakawa-Oiwake-cho, Sakyo-ku, Kyoto 606-8502, Japan}



\begin{abstract}

We present an extensive, panchromatic photometric (UV, Optical, and NIR) and low-resolution optical spectroscopic coverage of a Type IIP supernova SN~2018gj that occurred on the outskirts of the host galaxy NGC~6217. From the $V$--band light curve, we estimate the plateau length to be $\sim$\,70\,$\pm$\,2 d, placing it among the very few well-sampled short plateau supernovae (SNe). With $V$-band peak absolute magnitude  $\rm M_V\leq-17.0\pm0.1~mag$, it falls in the middle of the luminosity distribution of the Type II SNe. The colour evolution is typical to other Type II SNe except for an early elbow-like feature in the evolution of $V-R$ colour owing to its early transition from the plateau to the nebular phase. Using the expanding photospheric method, we present an independent estimate of the distance to SN~2018gj. We report the spectral evolution to be typical of a Type II SNe. However, we see a persistent blue shift in emission lines until the late nebular phase, not ordinarily observed in Type II SNe. The amount of radioactive nickel ($\rm ^{56}Ni$) yield in the explosion was estimated to be $\rm 0.026\pm0.007~M_\odot$. We infer from semi-analytical modelling, nebular spectrum, and 1-D hydrodynamical modelling that the probable progenitor was a red supergiant with a zero-age-main-sequence mass $\rm \leq 13~M_\odot$. In the simulated hydrodynamical model light curves, reproducing the early optical bolometric light curve required an additional radiation source, which could be the interaction with the proximal circumstellar matter (CSM).

\end{abstract}



\section{Introduction}
\label{sec:intro}
Massive stars ($\geq \rm 8~M_\odot$) reach their life-cycle termination in violent explosions termed as Core-Collapse Supernovae (CCSNe). On the basis of their observable properties, viz., spectral lines and light curves, CCSNe are further divided into various sub-classes. The absence of Hydrogen Balmer lines in the spectrum puts a supernova (SN) into the Type I class, whereas the presence of these features allows it to be classified as a Type II SN. Type I class is further subdivided with the presence or absence of \ion{He}{1} $\lambda~5876$ (Ib or Ic) \citep{1941Minkowski, 1997Filippenko}. Based on their light curve shape, Type IIs are further classified under Type IIP and IIL sub-classes. A plateau-like constant luminosity period of about 100~d in the light curve evolution means the SN is of the Type IIP (plateau) class. While a `linear' decline from maxima in the light curve denotes the SN to be of Type IIL (linear) class \citep{1979Barbon}. Although there are observed dissimilarities in the light curves of Type IIP and Type IIL classes, it is still unsettled if the two classes are intrinsically different. In many sample studies, it is noticed that the Type II sub-classes (IIP and IIL) form a continuous sequence \citep{2014Anderson,10.1093/mnras/stz1855, 2015ApJ...799..208Sc2,2018avinash,2019MNRAS.490.2799Dc3}. In contrast, few studies present subtle differences in these sub-classes both photometrically and spectroscopically \citep{2012ApJ...756L..30Ad1, 2014MNRAS.442..844Fd2, 2014MNRAS.445..554Fd3}. Recent studies with larger samples seem to favour the continuous population of these events \citep{2016AJ....151...33Gs1,2016ApJ...820...33Rs2, Gutierrez2017_TypeIISample}. Apart from these typical classes, there have been numerous instances where narrow emission spectral features are superimposed over usual broad emission features \citep{2012ApJ...756..173Si1, 2013A&A...555A..10Ti2, 2014MNRAS.438.1191Smith, 2020MNRAS.499..129Gi4, 2022MNRAS.513.3564Ri3} in the SN spectra. These events are interacting SNe and known as IIn supernovae \citep{1997Filippenko}.

With the advent of various night sky surveys, viz. Zwicky Transient Facility \citep[ZTF, ][]{2019ZTF}, Asteroid Terrestrial-impact Last Alert System \citep[ATLAS, ][]{2018PASP..130f4505T}, Gaia \citep{2016GAIA}, All-Sky Automated Survey for Supernovae \citep[ASAS-SN, ][]{2017PASP..129j4502K}, etc., we see an enormous number of discoveries and extensive follow-up related to SNe. From the numerous observational \citep{Elmhamdi2003Ni, 2014Anderson} and theoretical modelling studies \citep{2016Sukhbold, 2021Curtis}, it is well established that the average plateau length in type IIP SNe is around 100~d. However, there are cases when the plateau length is found to be longer \citep[$\geq 120~d$,][]{20062004et, 20092005cs, 20212020cxd} or shorter \citep[$\leq \rm 65~d$,][]{2021Hiramatsu,2022ApJ...930...34T} than the typical value. SNe with a longer plateau duration are mostly found to be low-luminosity (LL) IIP. A moderate range of progenitor masses \citep[$\rm 10-15~M_\odot$,][]{2014LLIIP} have been deduced by modelling them, and these masses corroborate with the direct detection of progenitors \citep{2009ARA&A..47...63S}. Nevertheless, Type IIP events with plateau lengths less than 70~d are rare in synthesised models and observations. The occurrence of short plateau events is very small ($\sim$4\% of all type IIP SNe) in binary population synthesis and single progenitor models \citep{2018PASA...35...49E, 2021Hiramatsu}.  The small number of well-studied short plateau objects is insufficient to constrain these rates quantitatively. Thus, any addition to the sample of short plateau events will further improve our understanding of Type II SNe.

The shorter plateau length in Type IIP SNe is usually explained by considerable stripping of the hydrogen envelope. The stripping of the outer hydrogen layer is possible in all mass ranges of red-supergiants (RSGs) via various mechanisms, viz. wind mass loss, presence of a secondary star, episodic mass losses, etc. Some theoretical works have shown high mass RSGs as progenitors of short plateau SNe \citep{2010MNRAS.408..827D, 2016Sukhbold, 2021Hiramatsu} as, with the usual single-star evolutionary scenario with ``typical" mass-loss, only the more massive stars are able to achieve stronger winds required to strip enough Hydrogen to cause a shorter plateau. However, there are reasons to believe otherwise, where it is possible to get a shorter plateau if the multiplicity \citep{2018PASA...35...49E} or extensive mass loss in lower mass RSG is considered \citep{2022ApJ...930...34T}. In a study by \citet{2021A&A...655A.105SJS}, the progenitor of the short plateau object SN~2020jfo was solely detected in the F814W band of the Hubble Space Telescope (HST), estimating its mass to range from 10 to 15~$\rm M_\odot$. The lack of detection of the progenitor in the bluer bands indicated the progenitor as a cool and red star. Further, the observational upper mass for Type IIP progenitors is $\lesssim 19~\rm M_\odot$ \citep{2017RSPTA.37560277V}, although the masses of RSGs observed in the Local Group galaxies have been found to range up to 25$\rm~M_\odot$ \citep{2009MNRAS.395.1409S, 2022MNRAS.515..897R}, leading to the missing mass ``RSG problem". Massive RSGs as progenitors for the short plateau could address the ``RSG problem" \citep{2021Hiramatsu}, although in general, there is no consensus with regards to the statistical significance of the RSG problem \citep{2013MNRAS.436..774E, 2020MNRAS.493.4945K, 2020MNRAS.496L.142D}.

With the high cadence of modern sky surveys, quite a few of the short plateau SNe are discovered soon after the explosion. Early detection of these events provides essential information about the late-stage evolution of the progenitor and its immediate surrounding. The `flash ionization' features seen in some of these events are interpreted as the presence of circumstellar material (CSM) in their immediate vicinity \citep{2014Natur.509..471G}. Flash ionization feature has been observed in about 18\% of the SNe II, at ages $<$ 5 days \citep{2016ApJ...818....3K18p}. With the increasing number of supernovae observed early enough, the fraction of supernovae showing flash ionization features is also increasing \citep[$>36\%$, ][]{2022arXiv221203313B36}, indicating the presence of CSM close to the explosion site is common in Type II SNe \citep{2018ApJ...858...15M}. Other observational features such as narrow emission lines, high-velocity absorption features in spectra, and enhanced luminosity in light curves \citep{2018MNRAS.476.1497Bullivant, 2019ApJ...882...68Singh, 2022MNRAS.509.2013Zhang} also reveal the presence of spatially confined CSM, most probably originating from enhanced mass loss from progenitor shortly prior to the explosion.

The evidence of CSM in Type IIP is usually seen early on with ionized lines, narrow emission lines, high-velocity absorption features in spectra, and enhanced luminosity in light curves \citep{2018MNRAS.476.1497Bullivant, 2019ApJ...882...68Singh, 2022MNRAS.509.2013Zhang}. Furthermore, it has been established that sometimes only a few of these features are present while others are missing altogether \citep{2019ApJ...885...43Andrews, 2021ApJ...906...56Dong}. However, alternate pathways regarding the initial rise times and enhanced luminosity are being studied theoretically. The detailed studies of the Hydrogen-rich layer in 3D models \citep{2022ApJ...933..164Goldberg} could also reveal the initial behaviour of the peak if other shreds of evidence are scarce.

This work presents detailed spectroscopic and photometric observations of SN~2018gj along with the analytical and hydrodynamical modelling to infer properties of probable progenitor. The paper has been divided as follows: Section~\ref{sec:data_reduction} briefly describes the data acquisition and reduction process. The analysis of the apparent and the bolometric light curve is given in Section~\ref{sec:light_curve_analysis}. This section also utilizes semi-analytical light curve modelling for approximate estimates of progenitor properties. The spectral evolution from the photospheric phase to the nebular phase is presented in Section~\ref{sec:spectra}. Further, in Section~\ref{sec:Progenitor}, we attempt to estimate the progenitor evolution history, its properties, explosion parameters, etc., using complete 1-D hydrodynamical modelling. We briefly discuss the implications of our work and aspects related to consistent blue-shifted emission lines in Section~\ref{sec:Discussion} and provide a conclusion in the subsequent section.

\begin{figure}
	 \resizebox{\hsize}{!}{\includegraphics{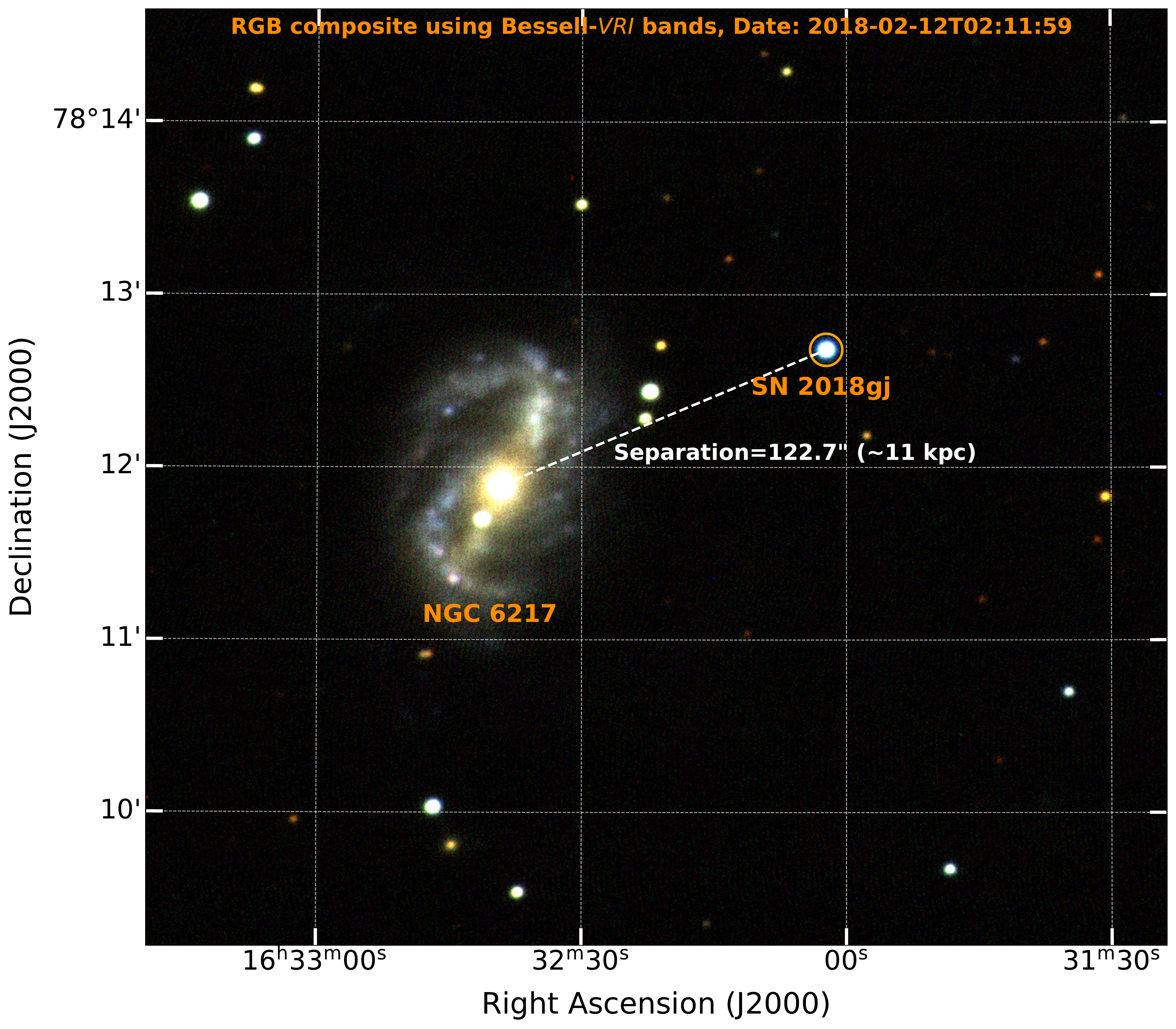}}
    \caption{Location of SN~2018gj in the host NGC 6217. The dashed violet line marks the separation between the host centre and SN. The image is an RGB colour composite utilising Bessell's $V$, $R$, and $I$ filters.}
    \label{fig:SN2018gjfield}
\end{figure}

\section{Data Acquisition and Processing}
\label{sec:data_reduction}

\begin{figure*}
	 \resizebox{\hsize}{!}{\includegraphics{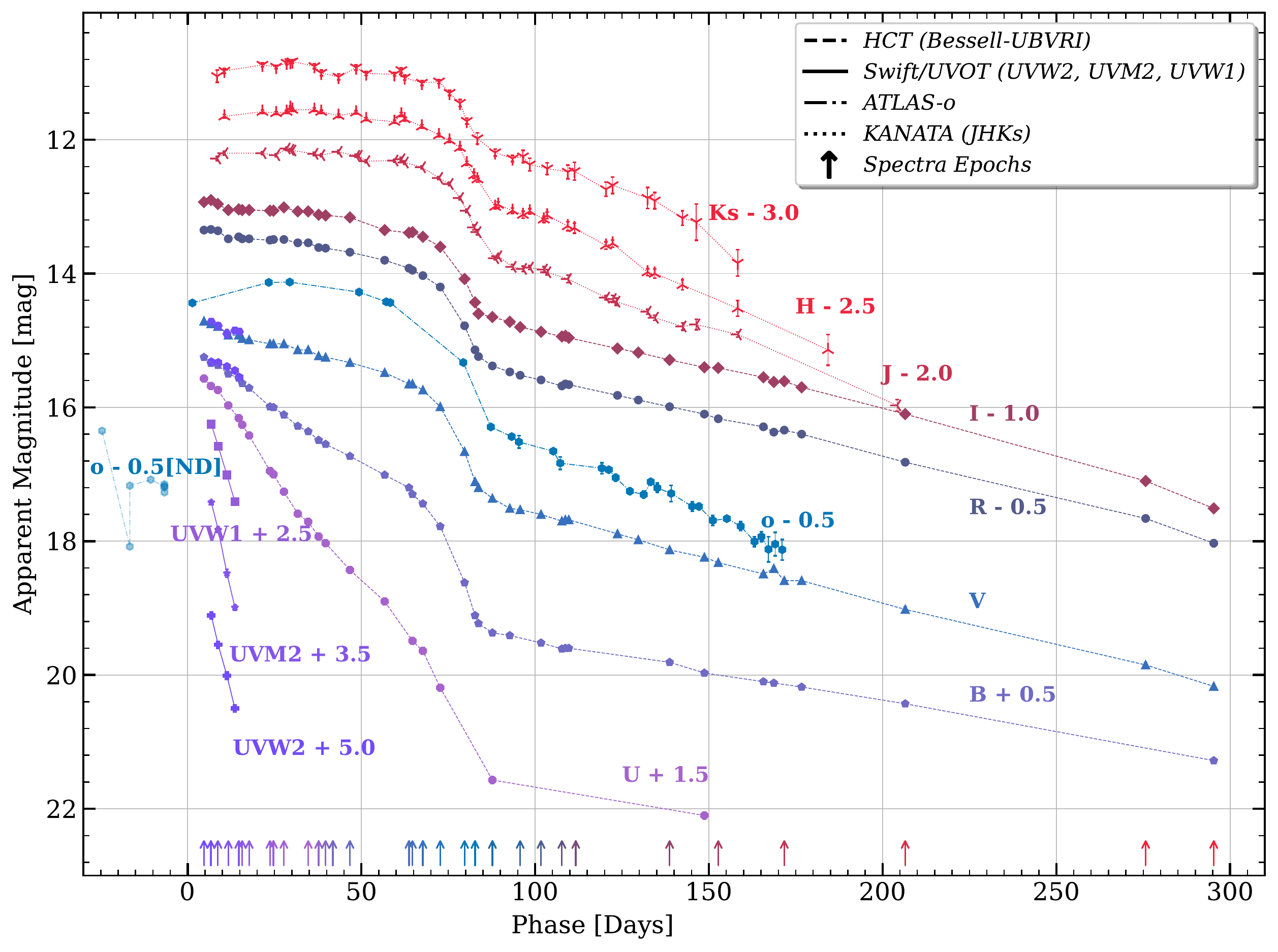}}
    \caption{Photometric data for SN~2018gj spanning $\sim$300~d post-discovery. Corresponding spectral epochs are marked along the abscissa. [Violet pentagon markers over-plotted on $V$ and $B$ bands are from Swift $UVV$ and $UVB$ bands, respectively]}
    \label{fig:light_curves}
\end{figure*}

The discovery of SN~2018gj, in the outskirts of the barred spiral galaxy NGC~6217 (about 122$\arcsec$ or $\rm \sim11~kpc$ away from the centre of the host galaxy) (Figure~\ref{fig:SN2018gjfield}) was reported on 2018 January 14 (2458132.91 JD) at  (J2000), $\rm RA, \alpha=16^\circ32\arcmin02\arcsec.40$ and $\rm Dec,\delta=+78^\circ12\arcmin41\arcsec.13$ \citep{2018TNS2018gj}. Immediately after the discovery, it was classified as a Type IIb SN with the possibility of it being a Type IIP SN  \citep{2018TNSCR..56....1B}. Later on, it was classified as a young Type II SN \citep{2018TNSCR2075....1K}. We started an extensive follow-up campaign in optical-NIR photometry and spectroscopy, which continued for about 300~d after the discovery. An imager cum spectrograph, the Himalaya Faint Object Spectrograph (HFOSC) mounted on the 2-m Himalayan Chandra Telescope (HCT) of the Indian Astronomical Observatory (IAO) \citep{2014Prabhu}, Hanle, India, was used for optical photometry and spectroscopy. It is equipped with a liquid nitrogen-cooled $2k\times4k$ pixels SITe CCD chip. With a pixel size of 15$\mu$m it provides a plate scale of $0.296\arcsec/pixel$. The gain and readout noise of the CCD are $\rm 1.22~e^-/ADU$ and $\rm 4.87~e^-$,  respectively. Near Infrared (NIR) data were acquired with  the Hiroshima Optical and Near-InfraRed Camera \citep[HONIR;][]{2014akitaya} installed at the 1.5-m Kanata Telescope operated by the Hiroshima Astrophysical Science Center (HASC) of Hiroshima University. The  NIR Arm has a HgCdTe VIRGO-2K array with $2k\times2k$ pixels (pixel size $\rm 20 \times 20~\mu m$, plate scale of $\rm 0.295\arcsec/pixel$) with $\rm 11.6~e^-/ADU$ and $\rm 24~e^-$ gain and readout noise, respectively. In addition to the science frames, we obtained several biases and twilight flat frames.

We pre-processed optical raw data using standard tasks in \texttt{IRAF} implemented using a custom pipeline \texttt{RedPipe} \citep{2021redpipe} built upon \texttt{PyRAF} to correct for bias, flat-field, and cosmic rays. The observed multiple frames were aligned and co-added in respective bands to improve the signal-to-noise ratio. Standard star fields PG 1633+009, PG 2213-006, and SA 110 from  \cite{Landolt1992} were observed  on four nights under photometric conditions to calibrate secondary standards in the SN field. We used DAOPHOT 3 \citep{Stetson1987} to perform point spread function (PSF) photometry on the standard  fields. The  average atmospheric extinction co-efficient  for the site \citet{Stalin2008} and  standard stars with a brightness range of $12.02\leq V \leq 16.25$ mag and colour range of $-0.22 \leq B-V \leq 2.53$~mag, were used  for calibration of secondary standards. As the SN was relatively isolated in its host galaxy,  the supernova magnitude was extracted using PSF photometry. The extracted magnitudes were then  calibrated differentially with respect to the secondary standard in the supernova field. The supernova magnitude in $UBVRI$ bands are listed in Appendix~\ref{appendix:data}. The NIR data was also reduced using the standard IRAF tasks. The secondary stars for $J$, $H$ and $Ks$ bands were calibrated using the magnitudes provided by the 2MASS catalogue \citep{2006AJ....131.1163S}. The supernova magnitude in $JHKs$ bands are listed in Appendix~\ref{appendix:data}.

Further, we supplemented our photometry data using public archive images from the UltraViolet/Optical Telescope \citep[][$UVOT$]{2005roming} on board Neil Gehrels {\it Swift} Observatory \citep{2004ApJ...611.1005G} in the $UVW2,\, UVM2,\, UVW1,\, UVU,\, UVB,$ and $UVV$ bands. The archival images were reduced using High Energy Astrophysics Software (HEASOFT, v6.27) package  with the latest calibration database for the $UVOT$ instrument, following the methods described in \citet{Poole2008} and \citet{Brown2009}. The SN magnitude was extracted using \texttt{UVOTSOURCE} task with an aperture size of $5\arcsec$ for the source and a similar aperture size to extract the background counts. The final $UVOT$ magnitudes were obtained in the Vega system and are tabulated in Table~\ref{tab:Swiftphotometry}. We also obtained photometry in ATLAS-$o$ band (AB-magnitude) from ATLAS forced photometry server \citep{2018PASP..130f4505T, 2020PASP..132h5002S}. 

The low-resolution spectroscopic  ($\sim$\,10~\AA) data was obtained with HFOSC using a setup consisting of $1.92\arcsec$ slit with grisms, Gr7 ($3800-6840$~\AA) and Gr8 ($5800-9350$~\AA) during  2018 January 14 (JD 2458132.5) to 2018 October 31 (JD 2458423.1). The log of spectroscopic observations is given in Table~\ref{tab:HCTspec} and marked in Figure~\ref{fig:light_curves}. Spectra of arc lamps and spectrophotometric standards were also obtained for wavelength and flux calibration, respectively. Standard tasks in IRAF were used for spectral data reduction. The observed spectroscopic data were corrected for bias, and the 1-D spectra were extracted using the optimal extraction algorithm \citep{1986horne}. Wavelength calibration was performed using the dispersion solutions obtained utilizing arc lamp spectra. Night-sky emission lines were used to check the accuracy of wavelength calibration, and small shifts were applied wherever necessary. The instrumental response was corrected  using the observed spectro-photometric standards. The response curves obtained during nearby nights were used for those nights where standard star observations were unavailable. The flux-calibrated spectrum in both the grisms was combined to obtain a single flux-calibrated spectrum. The spectra were then scaled to the calibrated $UBVRI$ magnitudes to bring them to an absolute flux scale.

\section{Analysis}
\label{sec:analysis}

\subsection{Host Properties}
\label{subsec:host}
The preferred redshift (z) and distance (D) of NGC~6217 are $\rm 0.00454\pm0.00001$ and $\rm 19.61\pm1.37~Mpc$, respectively, and are referenced from NASA/IPAC Extragalactic Database (NED)\footnote{\href{http://ned.ipac.caltech.edu/}{http://ned.ipac.caltech.edu/}}. Other distance estimates exist with a great scatter ranging from 15 Mpc to 35 Mpc \citep{1984A&AS...56..381B, 1997A&A...326..915T}. The SN was located in the outskirts of the probable host NGC~6217 ($\rm 9\arcmin13\arcsec.8\ ``W"\ and\ 47\arcsec.4\ ``N"\ implying\ \sim 2\arcmin$ separation from the host's centre). To check the veracity of its association with NGC~6217, an independent estimate of the distance to the SN was made using the Expanding Photosphere Method (EPM) \citep{Kirshner1974_EPM, Schimdt1992_EPM, hamuy2001_1999em}. The detailed methodology and calculations are presented in Appendix~\ref{appendix:EPM}. The  dilution factors were adopted from \citet{hamuy2001_1999em}, \citet{DH2005_EPMdilution}, and \citet{Vogl2019_EPMDilution}. We found that the distances estimated using constrained explosion epochs and non-constrained explosion epochs varied as much as by 3~Mpc. The average distances using all three dilution factors are $\rm 15.7 \pm 1.7~Mpc$ (non-constrained explosion epoch) and $\rm 17.5 \pm 4.1~Mpc$ (constrained explosion epoch). Errors quoted are due to the scatter in the different measurements for three filter sets and three dilution factors. The distance of the SN estimated using EPM is in agreement  with the distances given in NED for NGC~6217 and establishes the association of the SN with NGC~6217.

From IRSA-Galactic Dust Reddening and Extinction map \citep{2011ApJ...737..103Sirsa},  the Galactic line of sight reddening  towards the SN location is given as $\rm E(B-V) = 0.0375 \pm 0.0002~mag$.   Weak \ion{Na}{1D} absorption was detected in the spectra at the redshift of the host galaxy with a pseudo-equivalent width (pEW) of $\rm 0.36~\AA$, averaged over the first five spectra. Using \citep[Equation 9, ][]{2012MNRAS.426.1465PNaiD} we find a host reddening value of $\rm E(B-V) = 0.04 \pm 0.02~mag$. Hence, throughout this work, we adopt a total line of sight reddening value $\rm E(B-V) \approx 0.08 \pm 0.02~mag$.

\subsection{Light Curve Analysis}
\label{sec:light_curve_analysis}
The last non-detection of SN~2018gj was on 2018 January 7.9 (JD~2458126.4) in the Gaia photometry, up to the limiting magnitude of $\sim$21.5 in G-Gaia filter (AB magnitude system), and was discovered on 2018 January 10.7 (JD 2458129.2). Using this last non-detection and the first detection of SN~2018gj, the explosion epoch is constrained as  2018 January 9.3 ($\sim$JD~2458127.8) $\pm$ 1.4~d. This explosion epoch has been used throughout this work, and all the phases are reported with respect to it. The panchromatic light curve evolution of SN~2018gj in UV, optical and NIR bands are presented in Figure~\ref{fig:light_curves}. UV light curves in $UVW2$, $UVM2\,$ and $UVW1$ bands span a period of $\sim$14~d post-explosion, NIR light curves span up to 180~d whereas the optical light curve extends until $\sim$297~d. 
In all the light curves, we find a clear transition from the slowly declining (almost constant) light curve phase to the radioactive nickel-powered phase towards the end. During this transition, in visual bands, we see a drop of $\rm \geq 1.5~mag$.

During the plateau phase, the light curves decline at different rates in different wavebands. In the $U$ filter, we observe the sharpest decline with $\rm 6.44\pm0.03~mag~(100~d)^{-1}$. As we move towards the redder wavelengths, we find the decline rate slows down to $\rm 3.39\pm0.03~mag~(100~d)^{-1}$ in $B$, $\rm 1.26\pm0.02~mag~(100~d)^{-1}$ in $V$, $\rm 0.79\pm0.02~mag~(100~d)^{-1}$ in $R$ and $\rm 0.64\pm0.02~mag~(100~d)^{-1}$ in $I$. The decline is even slower in the near-infrared wavelength regime with $\rm 0.37\pm0.02~mag~(100~d)^{-1}$ in $J$, but the decline rate increases slightly as we go toward redder bands with $\rm 0.42\pm0.06~mag~(100~d)^{-1}$ in $H$ and ultimately $\rm 0.50\pm0.07~mag~(100~d)^{-1}$ in $Ks$ band. We find the slowest decline to be in the $J$ filter. It is also noteworthy that in the radioactive decay tail phase, the decline rate trend reverses with the slowest decline observed in $B$ band $\rm 0.90\pm0.01~mag~(100~d)^{-1}$ and it is almost the same in $V$, $R$ and $I$ bands as $\rm 1.33\pm0.01~mag~(100~d)^{-1}$, $\rm 1.14\pm0.02~mag~(100~d)^{-1}$ and $\rm 1.26\pm0.02~mag~(100~d)^{-1}$ respectively. We find the late-phase light curve decline rates to be much higher in the near-infrared bands as $\rm 1.86\pm0.07~mag~(100~d)^{-1}$ in $J$, $\rm 2.58\pm0.11~mag~(100~d)^{-1}$ in $H$ and $\rm 2.12 \pm0.24~mag~(100~d)^{-1}$ in $Ks$ bands. During the late phase, the light curve in the $H$ band declined at the fastest rate.

\subsection{V-band Light Curve}
\label{subsec:Vband}

\begin{figure}
	 \resizebox{\hsize}{!}{\includegraphics{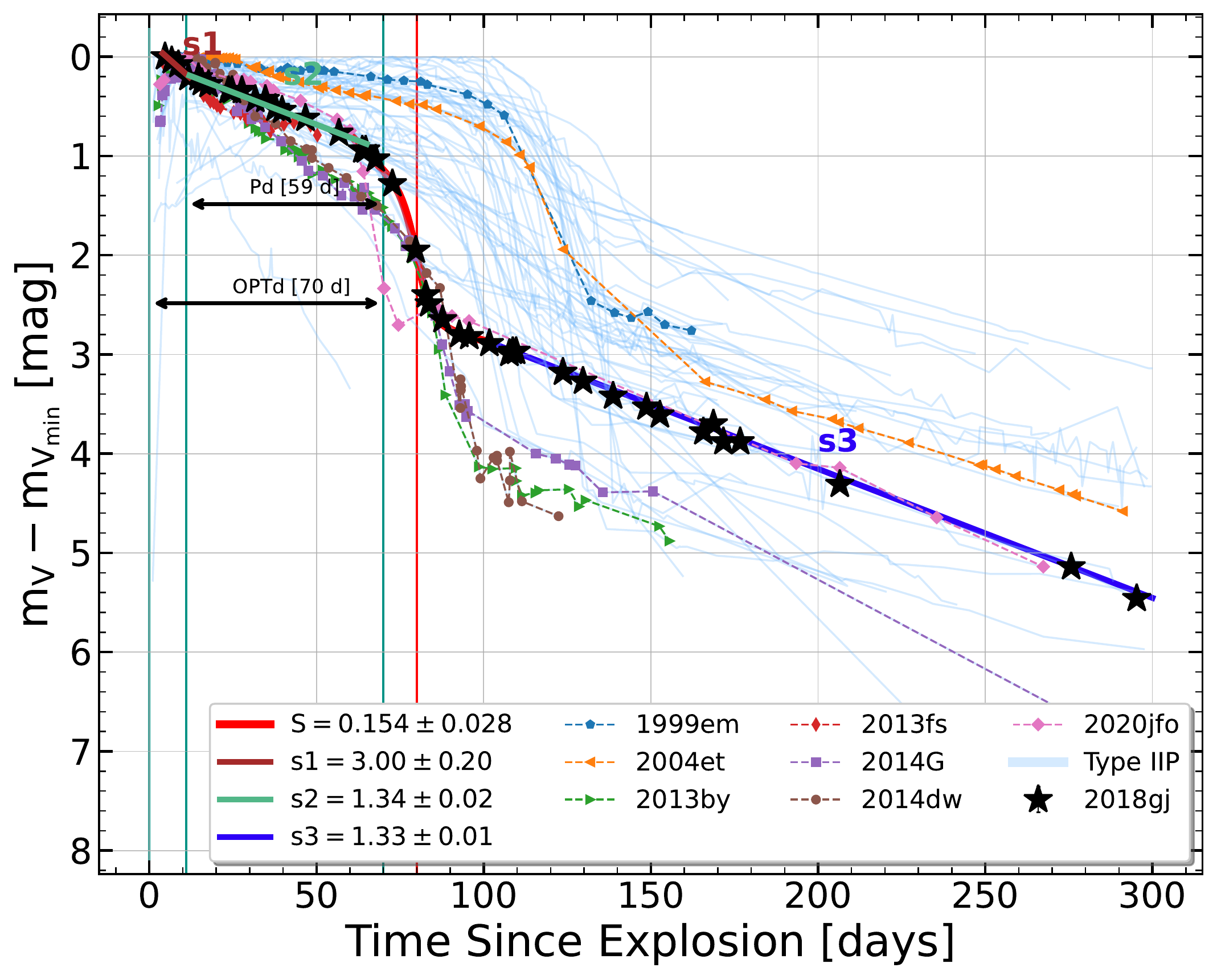}}
    \caption{$V$-band light curve evolution of SN~2018gj along with other Type II SNe. Continuous light blue lines are representative of a larger Type II sample from \citet{2014Anderson} and \citet{2014MNRAS.442..844Fd2}. Estimated light curve parameters for $V$-band viz. OPTd, Pd, s1, s2, and s3 are also shown. Supernovae data used in this plot are mentioned in Table~\ref{tab:SampleColor}}
    \label{fig:AbsoluteVnorm}
\end{figure}

After correcting for extinction, the apparent $V$-band magnitudes were transformed to absolute magnitudes using a distance modulus, $\rm \mu=31.46\pm0.15~mag$ (using the preferred distance of $\rm 19.61\pm1.37~Mpc$ as given in NED). Even though the initial rise in the bluer bands is missed, we see the light curve getting brighter during the first two observations in the $I$ band and, to a similar extent, in the $R$ band. A similar rise is observed in the NIR $J$ and $Ks$ bands. 

In the absence of the rising part of the V-band light curve, the peak absolute magnitude $\rm M_V$, could not be constrained well (Figure~\ref{fig:AbsoluteVnorm}). However, an  upper bound on the peak $\rm M_{V}  \lesssim -17.0\pm0.1$ mag can be set. The mean of maximum $\rm M_V$ value for a sample of 68 Type II SNe estimated by \citet{2014Anderson} is $\rm -16.74\pm1.01~mag$, which puts SN~2018gj towards the brighter end of Type II SNe. Furthermore, we observe a rapid decline in magnitude after $\rm +60~d$, corresponding to  a sharp transition from the plateau phase to the nebular phase. Using the functional form given in \citet{Elmhamdi2003Ni}, we could find the transition time at $\rm +79\pm2~d$ and a plateau length (OPTd) of $\rm \approx 70\pm3~d$, placing SN~2018gj in the shorter plateau group of Type IIP SNe. Following the phase definitions given in \citet{2014Anderson}, we estimated  $V$-band light curve parameters for SN~2018gj. We find s1, s2, and s3 to be  $\rm 3.00\pm0.20~mag~(100~d)^{-1}$,  $\rm 1.34\pm0.02~mag~(100~d)^{-1}$, and  $\rm 1.33\pm0.01~mag~(100~d)^{-1}$, respectively. The $s1$ and $s2$ decline rates are quite similar to the average values obtained from the Type II sample, which are $\rm 2.65\pm1.50~mag~(100~d)^{-1}$, and $\rm 1.27\pm0.93~mag~(100~d)^{-1}$, respectively. However, the decline in $s3$ is slower than the average Type II SNe $s3$ decline rate of $\rm 1.47\pm0.82~mag~(100~d)^{-1}$.


\subsection{Colours}

\begin{figure}
	 \resizebox{\hsize}{!}{\includegraphics{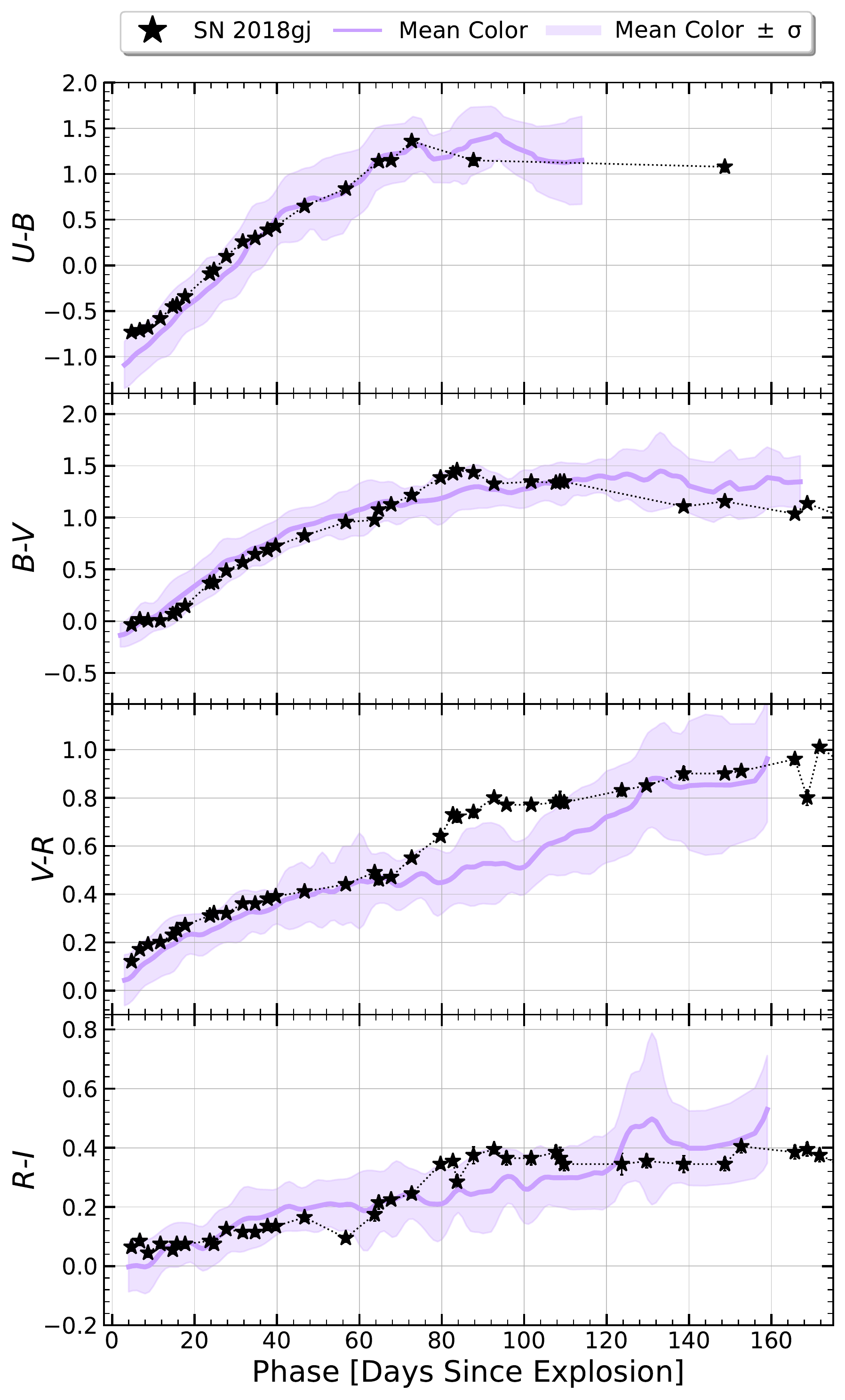}}
    \caption{Mean colour evolution of Type II SNe along with the colour evolution of SN~2018gj for different bands are shown. The shaded region colour with a solid line represents the mean colours from a larger Type IIP sample with $1\,\sigma$ scatter from the mean value. Sources of data have been referenced in the Table~\ref{tab:SampleColor}.}
    \label{fig:colors}
\end{figure}

In Figure~\ref{fig:colors}, the $U-B$, $B-V$, $V-R$, and $R-I$ colour evolution is shown. To compare SN~2018gj colours with other Type IIP SNe,  a mean colour curve from a sample of 44 Type IIP SNe,  available in the literature, is created (For reference, see Table~\ref{tab:SampleColor}). We do not consider any epoch on which the number of available data points is less than five. We apply extinction correction to all the respective individual band photometry using \citet{Cardelli} with $\rm R_V=3.1$.  Further, Gaussian smoothing is applied using \texttt{scipy.ndimage.gaussian\_filter1d}. The resultant mean colours, from the sample, along with $1\,\sigma$ scatter are over-plotted with that of SN~2018gj (see Figure~\ref{fig:colors}). The colour evolution of SN~2018gj  predominantly follows the typical Type IIP  SNe behaviour with slight deviations in early $U-B$, late $B-V$, and $V-R$ during the transition phase. The initial $U-B$ colour ($\rm <20~d$) for SN~2018gj is redder than the average $U-B$ value for Type IIP SNe whereas the $B-V$ colour evolution  of SN~2018gj starts to deviate after $\rm +110~d$ and becomes bluer than the average sample values. Further, we observe a slightly redder `elbow' kind of feature in $V-R$ mean colour values around $\rm +100~d$ for the sample, which could signify a mean plateau length duration of $\rm 100~d$ for the sample. In comparison, this break in $V-R$ colour evolution continuity is quite significant in SN~2018gj and is observed at $\rm +70~d$, which later evolves along with the mean colour evolution for the sample. The $R-I$ colour evolution SN~2018gj is typical of Type IIP SNe.    

\subsection{Bolometric Light Curve}

\label{subsec:bolom}
\begin{figure}
	 \resizebox{\hsize}{!}{\includegraphics{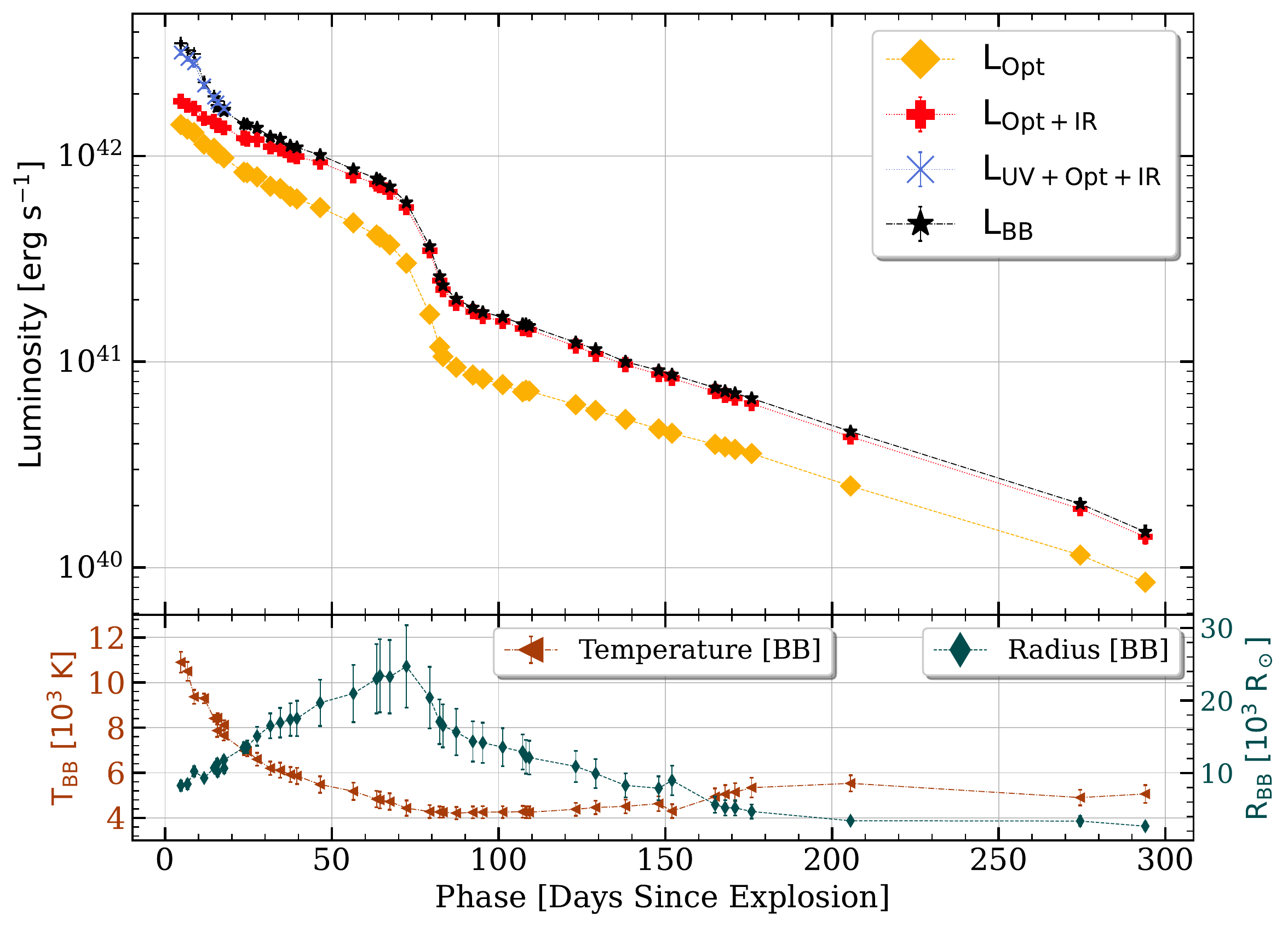}}
    \caption{Pseudo-bolometric and bolometric light curves for SN~2018gj obtained using multiband photometry are shown. The second plot at the bottom shows the temperature and radius evolution obtained using blackbody fits from the SEDs.}
    \label{fig:superbol_bol}
\end{figure}

The multi-broadband photometry is used to obtain the bolometric light curve of SN~2018gj, using the widely employed \texttt{SuperBol} \citep{mnicholl_2018_2155821} code. The code computes pseudobolometric/bolometric curves by integrating the flux over observed bands. Further, a complete bolometric curve is estimated using blackbody extrapolations, additionally providing information about the evolution of blackbody temperature and radius. The zero points used to convert magnitudes to fluxes are obtained from \citet{bessell} for UBVRIJHK of the Johnson-Cousins-Glass system, \citet{2018PASP..130f4505T} for the ATLAS filters. Zero points for other filters are obtained from SVO Filter Profile Service \citep{SVO1, SVO2}. To accommodate the missing epochs, the light curves were linearly interpolated, and if needed, the extrapolation was achieved using constant colour with respect to the well-sampled reference band. These objectives are utilized using  various tasks in \texttt{scipy}.

We estimate three different pseudobolometric/bolometric light curves. With only optical bands a pesudobolometric light curve ($\rm L_{Opt}$) is generated. Secondly, we include NIR data with optical and obtain OIR bolometric light curve (see Figure~\ref{fig:superbol_bol}). As the  UV data is not available throughout, we include UV data for the initial few days and estimate the bolometric light curve ($\rm L_{UV+Opt+IR}$). We find that using UVOIR data, the estimated bolometric light curve very closely traces the blackbody corrected estimate to the observed light curve ($\rm L_{BB}$). For further analysis, we use the UVOIR observed bolometric light curve.

We missed the early detection and rise, and therefore cannot constrain the peak in any of the bands. Hence, we only report the maximum value in the pseudobolometric/bolometric light curves. For the optical bolometric light curve, the peak value obtained is $\rm 1.42\pm0.06\times10^{42}~erg~s^{-1}$ and if we include the NIR and UV contributions, the values obtained are $\rm 1.84\pm0.06\times10^{42}~erg~s^{-1}$ and $\rm 3.18\pm0.08\times10^{42}~erg~s^{-1}$, respectively. 
We observe that during the initial phase of $\sim$\,5-15~d, the NIR contribution to the pseudobolometric light curve is only $\sim 25\%$. It sustains a maximum value of around $\rm \sim 50\%$ after the transition phase from 80~d to 110~d. The NIR contribution remains significant during the nebular phase also, with an average value of $\rm \sim 43\pm2\%$, which is similar to the values estimated for other SNe \citep{Patat_1998bw, Elmhamdi2003Ni}.

\subsection{Radioactive \texorpdfstring{$\rm ^{56}Ni$}{}}
\begin{figure}
	 \resizebox{\hsize}{!}{\includegraphics{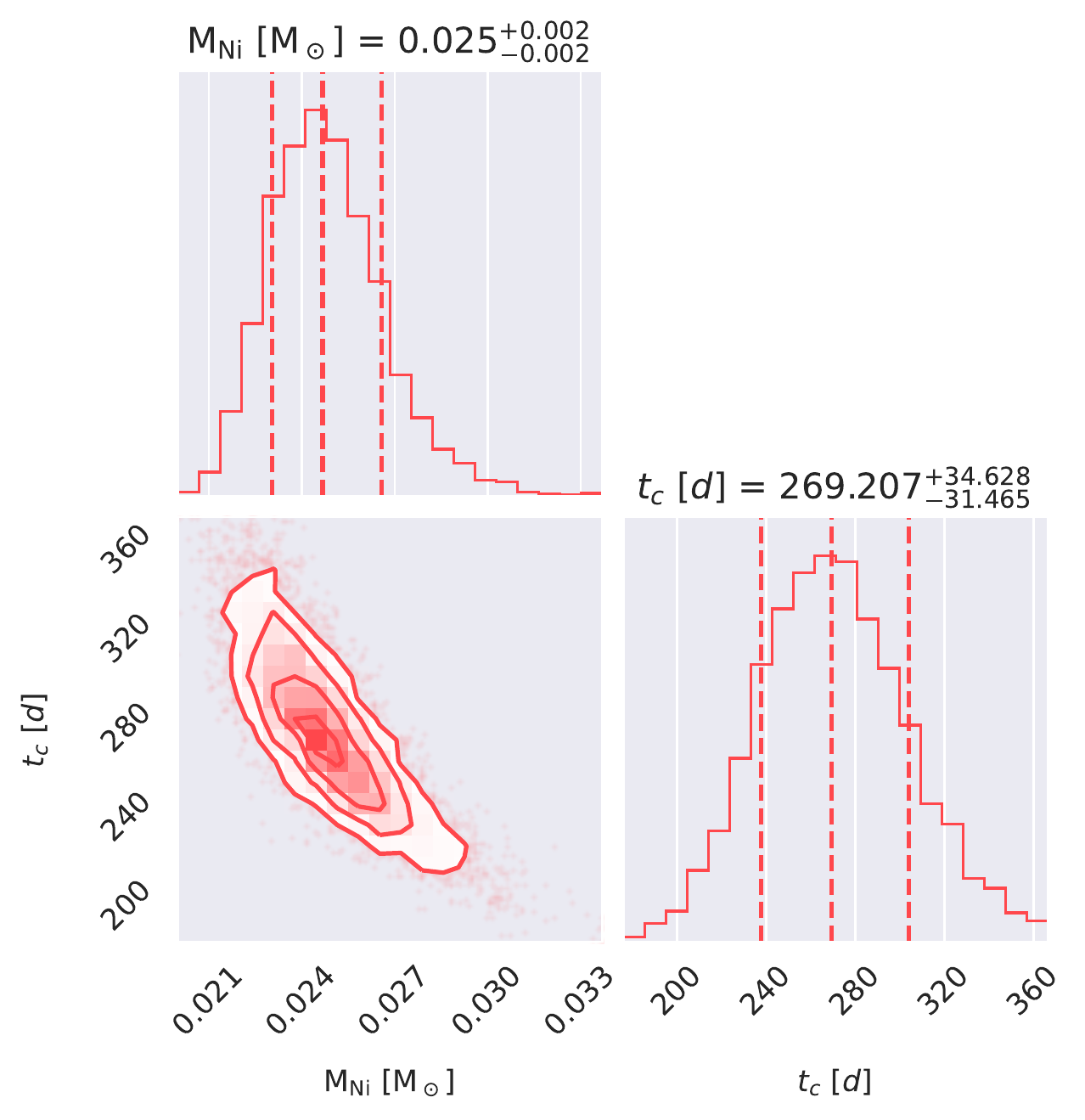}}
    \caption{Posterior plot for nickel mass and characteristic time estimates for SN~2018gj using Equation~\ref{eq:Nitc}.}
    \label{fig:NiTc}
\end{figure}

The $\rm ^{56}Ni$ decay chain primarily dominates the late-time light curve evolution of Type II SNe. It is the primary source of energy during the nebular phase of Type II SNe. We used various methods to estimate the mass of synthesized $\rm ^{56}Ni$. We compared the bolometric luminosity of SN~2018gj in the nebular phase with the bolometric luminosity of SN~1987A. The mass of $\rm ^{56}Ni$ in  SN~1987A is very well constrained using multiband photometry and hydrodynamical modelling and can be utilized to estimate mass of $\rm ^{56}Ni$ in SN~2018gj. We compare the bolometric luminosity with the values obtained for SN~1987A at similar epochs and use Equation~\ref{eq:Nickel1987A} to get an estimate on $\rm ^{56}Ni$ mass.

\begin{equation}
\label{eq:Nickel1987A}
    M_{Ni}(SN~2018gj) \approx M_{Ni}(SN~1987A)\times \frac{L_{1987A}(t)}{L_{2018gj}(t)}
\end{equation}
From the late time light curve ($\rm >110~d$) we estimate the mass of $\rm ^{56}Ni$ to be $\rm M_{Ni} = 0.024\pm0.004~M_\odot$. $\rm ^{56}Ni$ and characteristic timescale are also estimated using the Equation~\ref{eq:Nitc} and \texttt{scipy.minimize} and \texttt{emcee} packages.
\begin{equation}
    \label{eq:Nitc}
    L_{obs} = L_0\times M_{Ni}\times\left [ e^{-(\frac{t-t_0}{t_{Co}})}-e^{-(\frac{t-t_0}{t_{Ni}})}\right ]\times \left ( 1-e^{(-\frac{t_c^2}{(t-t_0)^2})}\right)
\end{equation}
The posterior distribution for the fits is shown in Figure~\ref{fig:NiTc}. We obtained  $\rm ^{56}Ni$ mass and characteristic time of $\rm 0.025\pm0.002~M_\odot$ and $\rm 269\pm33~d$, respectively. 

The steepness parameter (S) described in \citet{Elmhamdi2003Ni} could also be used to estimate the mass of $\rm ^{56}Ni$ synthesized in the explosion. Applying the refined steepness relation from \citet{2018avinash}, we get a steepness parameter, $\rm S=0.154\pm0.028$, which translates to $\rm M_{Ni}=0.028\pm0.005$. The estimated $\rm ^{56}Ni$ mass with different techniques are in good agreement with each other, with an average value of  $\rm M_{Ni} = 0.026\pm0.007~M_\odot$.

\subsection{Two-Component Analytical Light Curve Model}

\begin{figure}
	 \resizebox{\hsize}{!}{\includegraphics{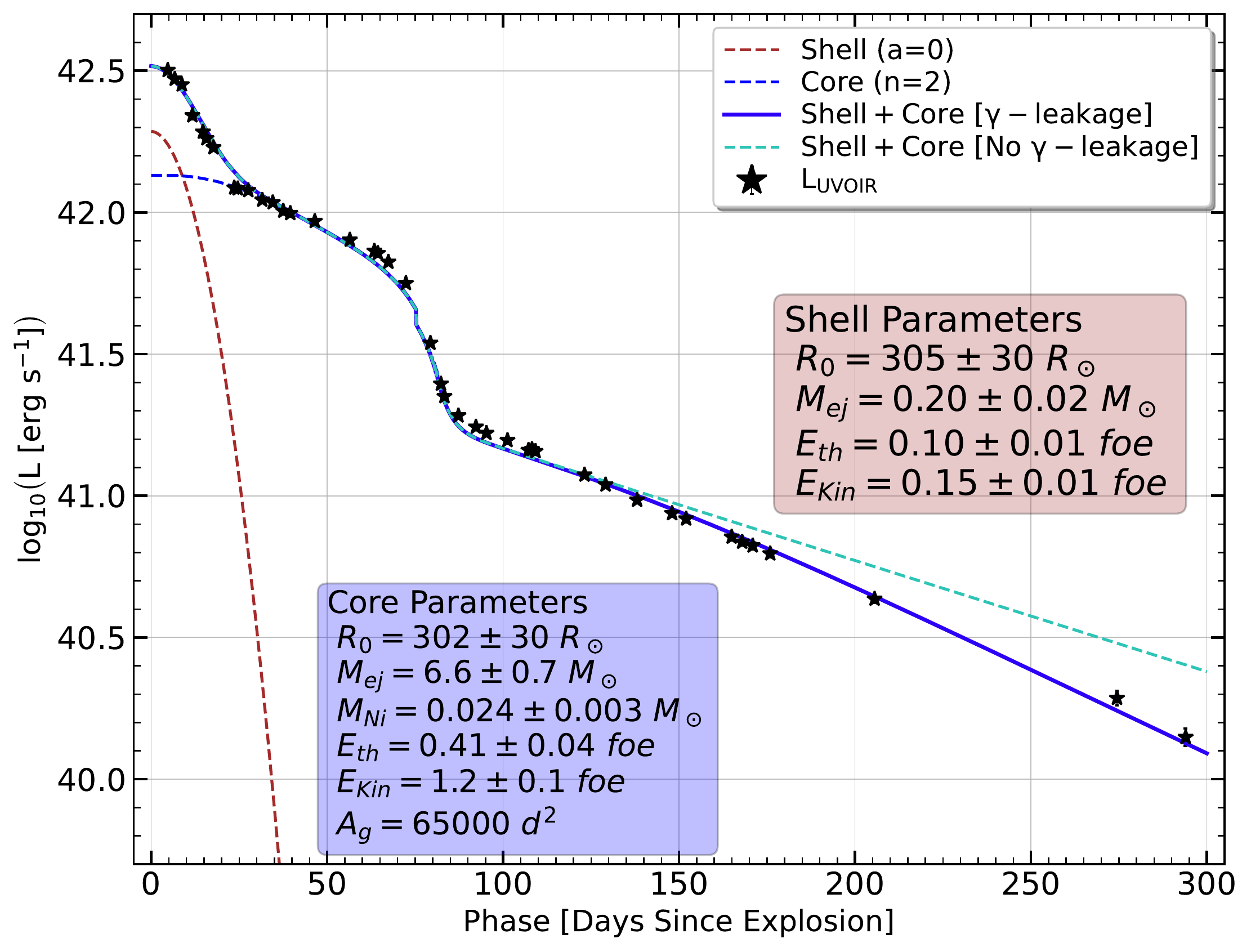}}
    \caption{Semi-Analytical model fitting for SN~2018gj using two-component model as described in \citet{NagyVinko2016}. The contributions from the shell and the core are also shown independently. In the inset, best-fitting parameters are listed for reference. The evolution of the light curve without $\gamma-$ray leakage is shown by the cyan dashed line.}
    \label{fig:Nagy_fit}
\end{figure}

\citet{NagyVinko2016} formulated a two-component progenitor model to  fit the observed bolometric light curves of Type IIP SNe semi-analytically. This formulation is based on the seminal work by \cite{ArnettFu1989} and subsequent modifications by \citet{Popov1993, BlinnikovPopov1993, Nagy2014}. It could be utilized to get approximate estimates on ejecta mass ($\rm M_{ej}$), progenitor radius ($\rm R_0$), total energy ($\rm E_{tot}$), and synthesized $\rm ^{56}Ni$ mass ($\rm M_{Ni}$). The formulation divides the homologously expanding and spherically symmetric SN ejecta into two components: an inner faction with a flat (constant) density configuration; an outer region with power law or exponential density profile \citep{Nagy2014}. Both these spherically symmetric components have different masses, radii, energies, and densities  but a common centre. The outer region is an extended envelope \citep{NagyVinko2016}. Contribution to bolometric luminosity ($\rm L_{bol}$) is primarily by energy released due to recombination ($\rm L_{rec}$) and radioactive decay ($\rm L_{Ni}$) of $\rm ^{56}Ni$. We use the UVOIR bolometric luminosity to approximate the semi-analytical models. The best fitting model is shown in Figure~\ref{fig:Nagy_fit}, and the obtained parameters are presented in Table~\ref{tab:nagy}. For the shell component we found, $\rm M_{ej_s}=0.2~M_\odot$ confined within a radius of $2.12\times10^{13}~cm$. We find a similar radius value for the core as well ($\rm \sim 2.10\times10^{13}~cm$) with an ejecta mass, $\rm M_{ej_c}=6.6~M_\odot$. The outer envelope appears not far-extended, and the density is slightly higher ($\rm \sim1.0\times10^{-8}~g~cm^{-3}$) as obtained for other Type IIPs in \citet{NagyVinko2016}. For comparison, the radii obtained are between the values obtained for SN~2005cs ($\rm R_{shell}=2.0\times10^{13}~cm$ and $\rm R_{core}=1.2\times10^{13}~cm$) and SN~2012aw ($\rm R_{shell}=4.5\times10^{13}~cm$ and $\rm R_{core}=3.0\times10^{13}~cm$). The shell densities obtained for both the cases are $\rm 1.8\times10^{-8}~g~cm^{-3}$ and $\rm 5.2\times10^{-9}~g~cm^{-3}$ respectively. The values obtained for SN~2018gj are within similar ranges for other Type IIP SNe with a normal plateau duration. From the semi-analytical modelling, we get a total ejecta mass, $\rm M_{ej}\approx6.8\pm0.7~M_\odot$, radius, R$\approx305\pm30~R_\odot$, and $1.9\pm0.2$ foe as the total energy released after the explosion. 

During the nebular phase, the light curve decline rate of SN~2018gj is $1.34\pm0.02$~mag\,100$\rm\,d^{-1}$, much faster than $0.98\pm0.02$~mag\,100$\rm\,d^{-1}$, the decay rate of $\rm ^{56}Co$ to $\rm ^{56}Fe$ with full $\gamma-$ray trapping. The faster decline of the late-phase light curve indicates that the leakage of $\gamma$-rays is significant in SN~2018gj. The effect of $\gamma-$ray leakage on the late time light curve  could be introduced using the $\rm A_g$ parameter. This parameter is the effectiveness of  $\gamma-$ray trapping \citep{2012Chatzopoulos} whereas in luminosity equation, it could be shown as $\rm L_{bol}=L_{Ni}(1-\exp{(-A_g/t^2)}) + L_{rec}$. Physically, it is related to the characteristic time scale ($\rm T_0$) of the $\gamma-$rays as $\rm A_g = T_0^2$. The late phase light curve,  powered by the radioactive decay, is fit by an $\rm A_g=65000~d^2$ and mass of synthesized $\rm ^{56}Ni$,  ($\rm M_{Ni})=0.025~M_\odot$. The corresponding $\rm T_0$ value is 255~d, similar to the value obtained in Section~\ref{subsec:bolom}. The $\rm M_{Ni}$ estimated here corroborates our previous estimates in Section~\ref{subsec:bolom}. Further, the correlation between ejecta mass and opacity \citep[correlation coefficient, r $= 0.984$, ][]{NagyVinko2016} makes it insubstantial to comment on the possible progenitor mass with certainty up to two orders of magnitudes. If we consider a proto-neutron star core of mass $\rm \sim 1.5~M_\odot$, nominal mass loss due to winds, and the estimated ejecta mass, we could constrain the lower limit of progenitor mass, which is $\rm \geq 10-11~M_\odot$, an estimate very similar to another short plateau object SN~2020jfo \citep{2022ApJ...930...34T}, which had a very similar light curve shape, but a shorter plateau by $\sim$\,10 days.

\begin{table}
\centering
\caption{Parameters for best-fitting two-component model} 
\begin{tabular}{l|c|c} \hline
   Parameters$^*$          & Shell         & Core        \\        
\hline
Ejecta Mass, $\rm M_{ej}$ ($\rm M_\odot$) & 0.20& 6.60 \\
Radius, R ($\rm 10^{13}~cm$)& 2.12 & 2.10 \\
Thermal Energy, $\rm E_{th}$ ($\rm 10^{51}~erg$)& 0.10 & 0.41 \\
Kinetic Energy, $\rm E_{kin}$ ($\rm 10^{51}~erg$)& 0.15 & 1.2 \\
Expansion Velocity, $\rm v_{exp}$ ($\rm 1000~km~s^{-1}$)& 13.0 & 5.5 \\
Opacity, $\rm \kappa$ ($\rm cm^2~g^{-1}$) & 0.4 & 0.2 \\
\hline
\end{tabular}\\
$^*$ $\rm T_{rec}\approx6000~K$, $\rm A_g=6500~d^2$ and $\rm M_{Ni}=0.024~M_\odot$
\label{tab:nagy}
\end{table}

\begin{figure*}
	 \resizebox{0.9\hsize}{!}{\includegraphics{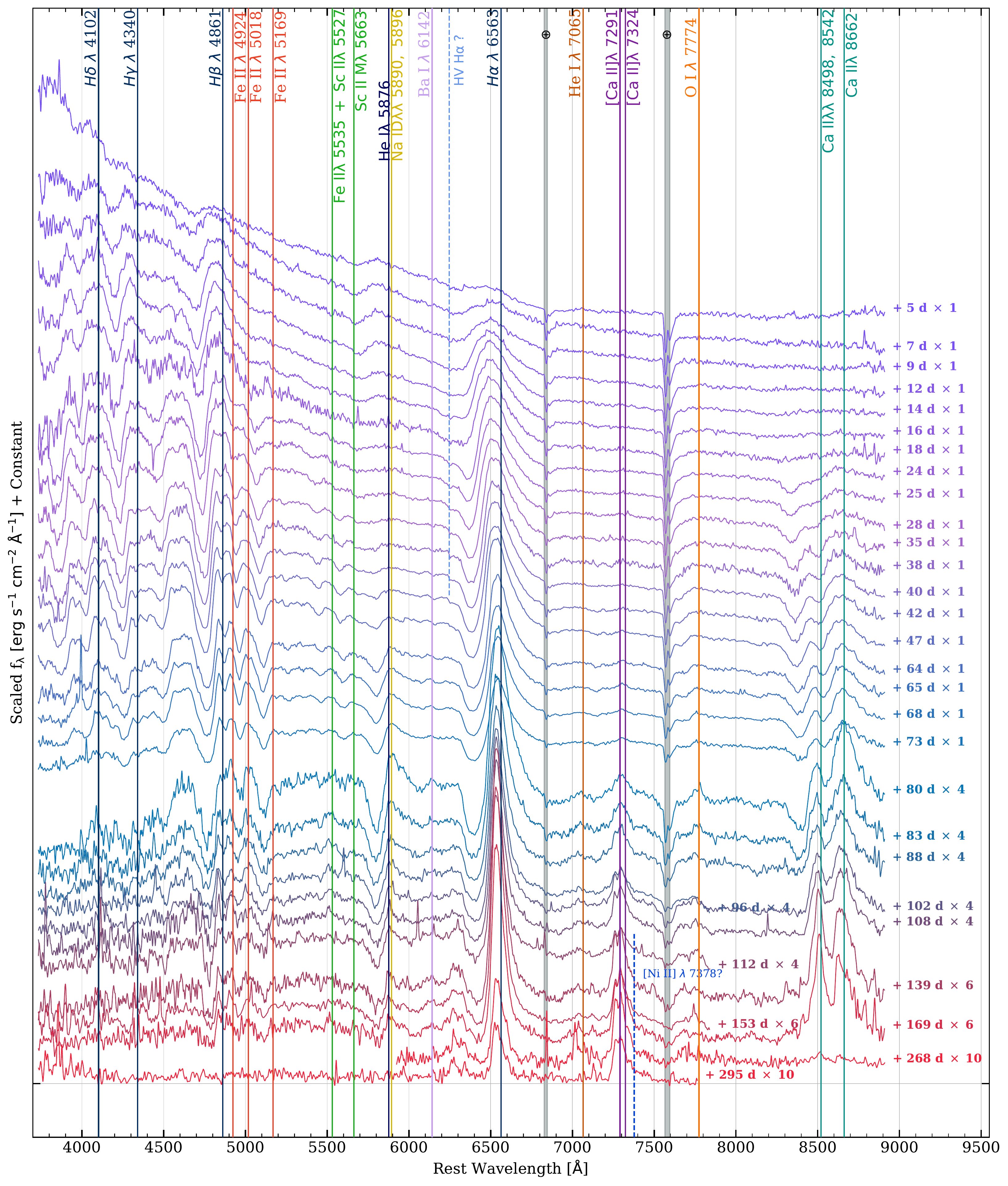}}
    \caption{Spectral time series for SN~2018gj containing 31 epochs spanning 295 d post-explosion. All spectra have been calibrated with photometry for absolute flux and corrected for host redshift. Some of the prominent spectral lines have been marked for clarity. (The data used to create this figure are available.)}
    \label{fig:spectra_evolution}
\end{figure*}

\section{Spectral Analysis}
\label{sec:spectra}
Apart from SN classification, detailed spectral studies provides insight into the ejecta composition, asymmetries, dust formation, and explosive nucleosynthesis. In this section, we present a detailed optical spectroscopic analysis of SN~2018gj. The temporal evolution of spectra is presented in Figure~\ref{fig:spectra_evolution}, marked with some well-identified hydrogen and metal features. The spectral sequence is not corrected for telluric absorption lines. Further, all the spectra have been scaled with photometry for absolute flux calibration and corrected for the host redshift. We study the spectral evolution spanning 31 epochs over the photospheric phase to the nebular phase beginning $+5~d$.

\subsection{Photospheric/Plateau Phase Spectra}

\begin{figure}
	 \resizebox{\hsize}{!}{\includegraphics{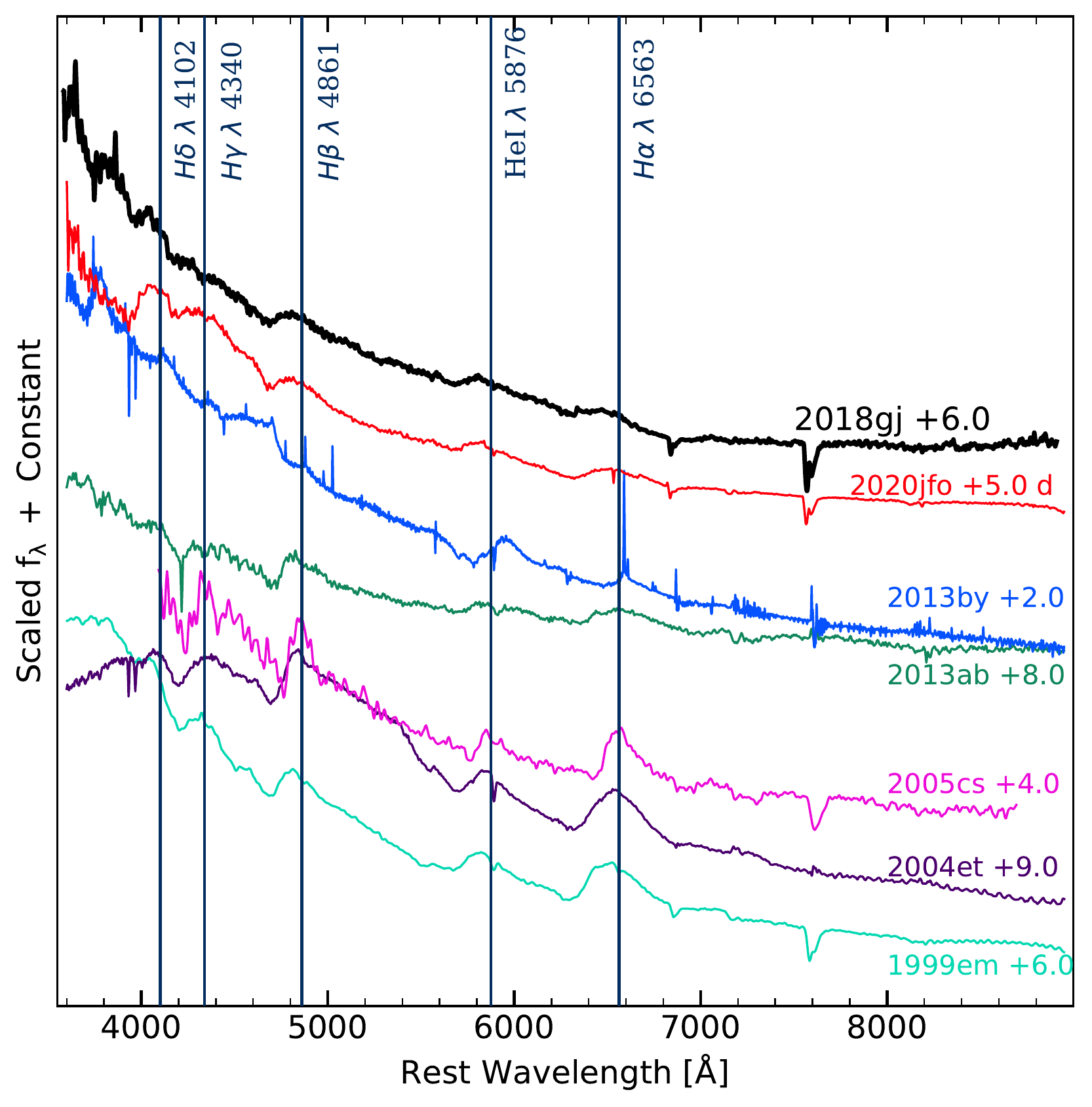}}
    \caption{SN~2018gj spectrum during maximum light. Spectra of other Type II SNe around the maximum is shown for comparison. The comparison sample is drawn from Table~\ref{tab:SampleColor}.}
    \label{fig:spectraComaprisonearly}
\end{figure}

The early part of spectra before or around the peak for typical Type II SNe is dominated by a featureless blue continuum along with a hint of formation of broad and discrete Hydrogen features, predominantly Balmer series ($\rm H\alpha~6563~\AA,\ H\beta~4861~\AA,\ H\gamma~4340~\AA\ and\ H\delta~4102~\AA$). The features show a typical P-Cygni profile due to the expansion of the ejecta. The early spectra of SN~2018gj show these features.  The \ion{He}{1}$\rm ~5876~\AA$ appears as early as $\rm+5~d$ and seen till $\rm+16~d$ where it gets blended with the \ion{Na}{1D}$\rm~5890,~5896~\AA$. The temperature of the ejecta estimated using SED (Section~\ref{subsec:bolom}) around this phase is about $\rm \gtrsim10000~K$, as the ejecta expands, it gradually cools down. With the ejecta cooling, metallic lines are seen, dominating the blue region of the spectra. All these metallic features show well-defined P-Cygni profiles. As the SN evolves, the absorption depth increases in strength and the  \ion{Fe}{2} multiplet  $\rm~4924,~5018,~5169~\AA$ lines are clearly seen at $+$24~d. The near-infrared region of the spectrum evolves with conspicuous \ion{Ca}{2} triplet ($\rm~8498,~8542~,~8662~\AA$) that is visible during the same phase and becomes prominent as it evolves further. Towards the end of the plateau around $\rm \sim+64~d$, the \ion{Na}{1D} line develops prominently.

The spectral evolution covers the transition from the plateau phase to the nebular phase very well. We obtained four spectra during the transition period from $\rm +72~d$ to $\rm +88~d$. We find increased flux in the redder side with the increased strength of \ion{Ca}{2} triplet. Similarly, other features become more prominent with an increase in their strengths. Apart from the increasing strength of earlier Hydrogen features, other lines, viz. \ion{Ba}{1}$\rm ~6142~\AA$, \ion{He}{1}$\rm~7065~\AA$, and \ion{O}{1}$\rm~7774~\AA$ develop and are observed clearly (Figure~\ref{fig:spectra_evolution}). This could be either due to the temperature change or because we can probe deep inside the ejecta as the hydrogen layer becomes transparent to the radiation from these parts. Nevertheless, from the SED fitting, we find the temperature fairly consistent within this phase. So this is primarily due to the decreased opacity of the hydrogen layer. We do not find other stark differences during the transition phase.  

\begin{figure}
	 \resizebox{\hsize}{!}{\includegraphics{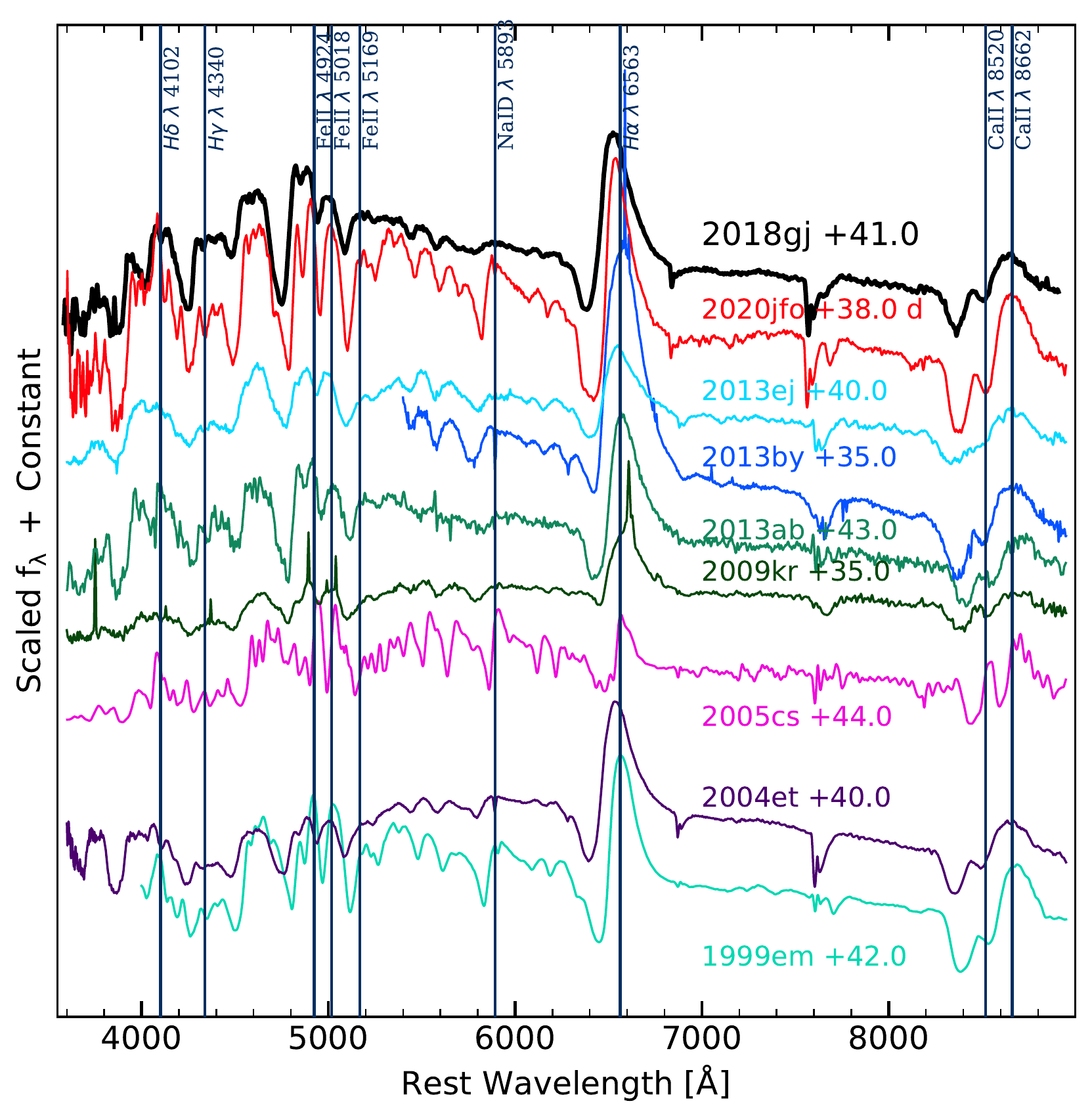}}
    \caption{Spectrum of SN~2018gj during plateau phase is shown in comparison with other Type II SNe. The comparison sample is drawn from Table~\ref{tab:SampleColor}.}
    \label{fig:spectraComaprisonplateau}
\end{figure}

When we compare the spectral features of SN~2018gj with other Type II SNe, we find that these features are fairly typical and are observed in all sorts of Type II SNe whether they show plateau or decline linearly both in the early phase (Figure~\ref{fig:spectraComaprisonearly}) as well as photospheric phase (Figure~\ref{fig:spectraComaprisonplateau}). The primary distinction is broadly the strength and spread of these features. During maximum light, the absorption trough of lines observed in SN~2018gj lies in-between other SNe used for comparison. For archetypal Type IIP SNe, SN~1999em \citep{hamuy2001_1999em, Leonard2002_1999em, ElmhamdiIR2003} and SN~2004et \citep{20062004et}, we find the strength of Balmar features is more prominent around the similar phase. However, even for normal Type IIP SNe, e.g., SN~2005cs \citep{Pastorello2006_2005cs, 2014MNRAS.442..844Fd2} and SN~2013ab \citep{Bose2015_2013ab} the strength could vary. Although the early spectrum of SN~2020jfo \citep{2022ApJ...930...34T}, another short plateau event, appears very similar to the spectrum of SN~2018gj except toward the shorter wavelengths, especially around H$\gamma$. There was the presence of ionized He in the early spectra of SN~2020jfo. Spectral comparison around mid plateau for SN~2018gj reveals the lack of metallic or fully developed features, which are more prominent in other SNe viz. SN~2020jfo, SN~2013ej, SN~2005cs, SN~2004et, and SN~1999em. Furthermore, it is observed that the hydrogen and other metallic features in SN~2018gj are weak as compared to normal Type II SNe but similar to SN~2009kr and SN~2013ab. SN~2013ab shows many similarities with SN~2018gj around the same phase.

\subsection{Ejecta Velocity}

\begin{figure}
	 \resizebox{\hsize}{!}{\includegraphics{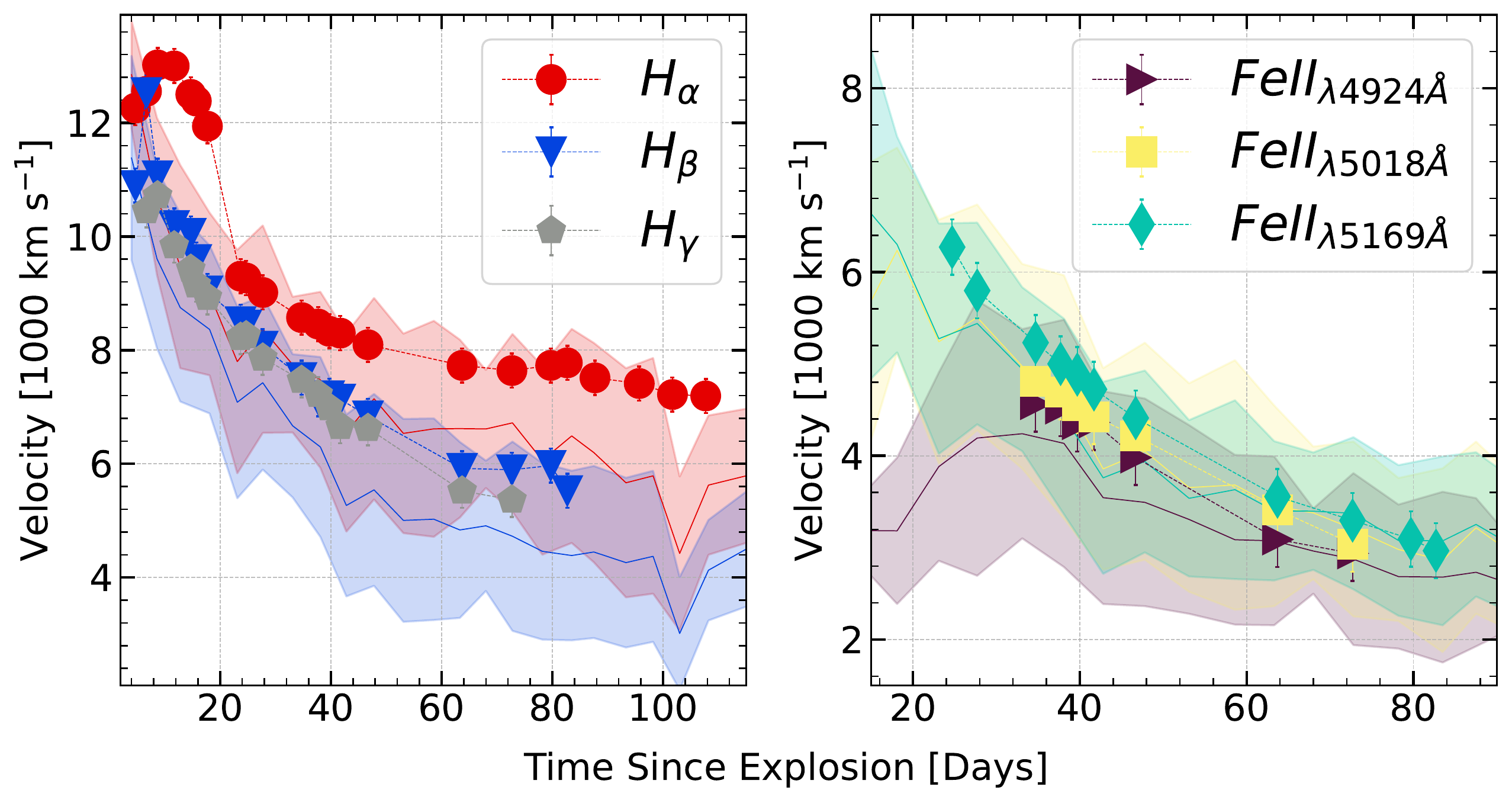}}
    \caption{Temporal velocity evolution of various lines identified in the spectra using the absorption minimum is shown here. The velocities have been compared with the sample from \citet{Gutierrez2017_TypeIISample} where the solid lines are the mean values from the sample and the shaded area around it in similar colour represents the 1$-\sigma$ scatter from the mean velocities.}
    \label{fig:vel_comp}
\end{figure}

In Figure~\ref{fig:vel_comp}, we show the expansion velocities estimated using the non-blended absorption minima of various species. The absorption minimum is estimated by fitting a Gaussian profile, and the expansion velocities are measured with respect to the rest frame wavelengths. A  peculiar velocity evolution of Hydrogen features is seen for the initial few days. It first rises and then declines. The rising part of the ejecta velocity has not been observed for other SNe. While the estimation of an initial lower velocity may indeed be true, the absorption features during this phase are very broad and associated with higher measurement uncertainty. A shallow absorption feature is seen around 6200\AA\ during the early phases (until $\sim 40$~d). It could be attributed to a high velocity (HV) H$\alpha$ feature \citep{2022A&A...660L...9D}, at a velocity of $\sim 15000$~km~s$^{-1}$. Figure~\ref{fig:vel_comp} also shows a comparison of the SN 2018gj velocities with the mean expansion velocities obtained from a large sample of Type II SNe  \citep{Gutierrez2017_TypeIISample}. The expansion velocities for SN~2018gj are towards the higher end of $1-\sigma$ scatter from the mean. It continues to follow this higher velocity trend even after transitioning from the plateau phase to the nebular phase. The expansion velocity inferred from \ion{Fe}{2} features  is found to be higher than the mean value initially; however, later, it follows a trend similar to the mean of the sample.

From various absorption features, we estimate that the layers of the ejecta are moving with velocities higher than $\rm10000~km~s^{-1}$. Although the temperature around a similar phase estimated for SN~1999em is similar to SN~2018gj, the H$\alpha$ velocity inferred was much higher ($\sim$ 16000~$\rm km~s~^{-1}$) \citep{ElmhamdiIR2003}. In the case of SN~2013ej \citep{Valenti2014_2013ej} and SN~2020jfo, the expansion velocities are around $\rm~13000~km~s^{-1}$ and comparable to that of SN~2018gj. As the ejecta evolves with time, it starts to slow down ($\rm < 9000~km~s^{-1}$ around +20~d)  and cool to a lower temperature ($\rm \leq 8000~K$ around +20~d). 

During the photospheric phase, the expansion velocities continue to follow the declining trend and reach  $\sim$8000~$\rm km~s^{-1}$ around +40~d. Afterwards, the decline is very slow and does not follow the average trend. The expansion velocities estimated using Hydrogen features are on the higher side for the Type II SNe. For SN~1999em around +40~d, typical temperatures are 5000--6000~K and H$\alpha$ velocity of about 6000$\rm~km~s^{-1}$ \citep{ElmhamdiIR2003}. In 2005cs velocities are much lower around +40~d and are estimated as $\rm \leq 4000~km~s^{-1}$ and $\rm 2000~km~s^{-1}$ for H$\alpha$ and metal lines, respectively \citep{Pastorello2006_2005cs}. Around similar phase H$\alpha$ and metal velocities for SN~2004et are 7500 $\rm km~s^{-1}$ and 4500 $\rm km~s^{-1}$, respectively. In case of SN~2009kr, SN~2013by and SN~2020jfo, H$\alpha$ velocities are $\rm \leq 7000~km~s^{-1}$  whereas for SN~2018gj it is close to the velocities estimated for SN~2013ab (8000~$\rm km~s^{-1}$) and SN~2013ej (8500~$\rm km~s^{-1}$).  

Similar observations are true for velocities estimated using metal lines. Typical expansion velocities around the plateau phase start at $\rm 6000~km~s^{-1}$ and slow down to $\rm \sim 3000~km~s^{-1}$ towards the end of the plateau phase.

\subsection{Nebular Phase Spectra}
When the receding photosphere reaches the base of the outer envelope, the plateau/photospheric phase ends, and the supernova transitions to the nebular phase. Continuous expansion reduces temperature, column density, and optical depth \citep{2017hsn..book..795J}. The ejecta is optically thin, and the inner ejecta layers are probed. Supernova in this phase is still optically bright, and the prominent energy source is the radioactive decay chain of \  $\rm ^{56}Ni$ synthesized in the explosion. The midpoint of transition happens around +80~d, and several metal lines originating from forbidden transitions, e.g. [\ion{Ca}{2}]$~7291,~7324~\AA$ start appearing. 
The strength of \ion{Na}{1D}~$~5893~\AA$ and \ion{Ca}{1} triplet keeps on increasing as the ejecta evolves with time. Other forbidden lines viz. [\ion{Fe}{2}]$~\rm~6118,~7155,~7172~\AA$, [\ion{O}{1}]$~\rm6300,~6364~\AA$ also start to appear in the spectra. Blue-ward of Ca triplet, we identify the \ion{O}{1}$~\rm7774~\AA$. The presence of H$\alpha$ continues during the nebular phase and  is the  dominant line  in the spectrum, although much narrower. 

\begin{figure}
	 \resizebox{\hsize}{!}{\includegraphics{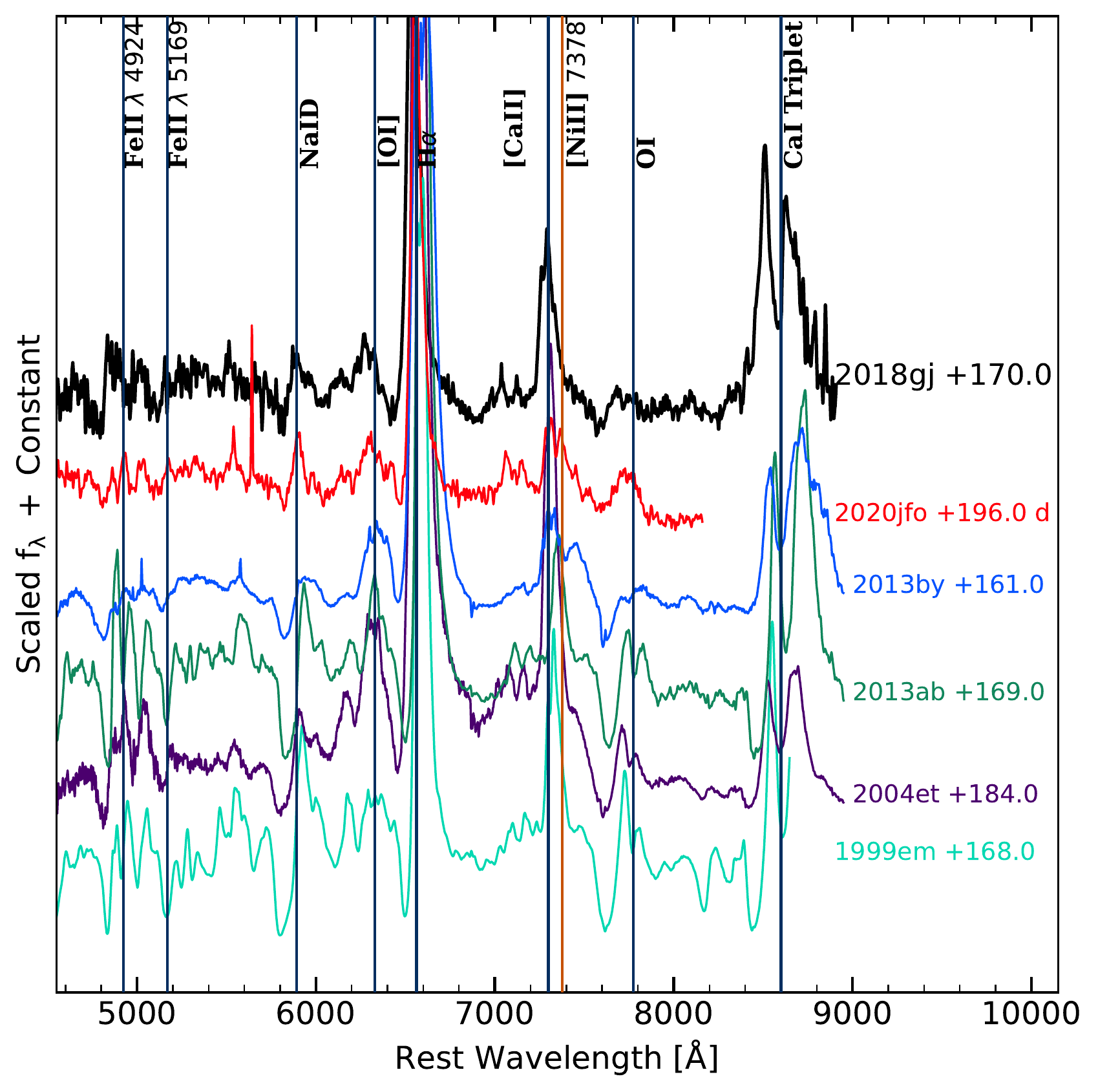}}
    \caption{Spectral comparison of SN2018gj with other Type II SNe around similar epochs during the nebular phase.}
    \label{fig:spectraComaprisonnebular}
\end{figure}

Figure~\ref{fig:spectraComaprisonnebular} shows the comparison of nebular phase spectra for SN~2018gj with several other Type II SNe. The \ion{O}{1}$~\rm7774~\AA$ line in SN~2018gj is found to be weakest in comparison to other Type II SNe except for SN~2013by. Apart from certain features common in Type II SNe, we find a hint of stable [\ion{Ni}{2}]~$\rm7378~\AA$ with emission feature having an intrinsic velocity similar to other emission features starting from $+112~d$ (Figure~\ref{fig:spectra_evolution}). This feature was observed in SN~2020jfo $\sim$  $\rm +196~d$ as presented in \citet{2022ApJ...930...34T}. However, it is quite possible that in SN~2020jfo, the stable Ni was present from an earlier epoch, but due to its proximity to the Sun, the first nebular spectrum could be  obtained $\sim$  $\rm +196~d$. This feature is very prominent in SN~2020jfo but weak in SN~2018gj.

\subsection{Blue-shifted Emission From Photospheric Phase To Nebular Phase}
\label{subsec:blushift}
\begin{figure}
	 \resizebox{\hsize}{!}{\includegraphics{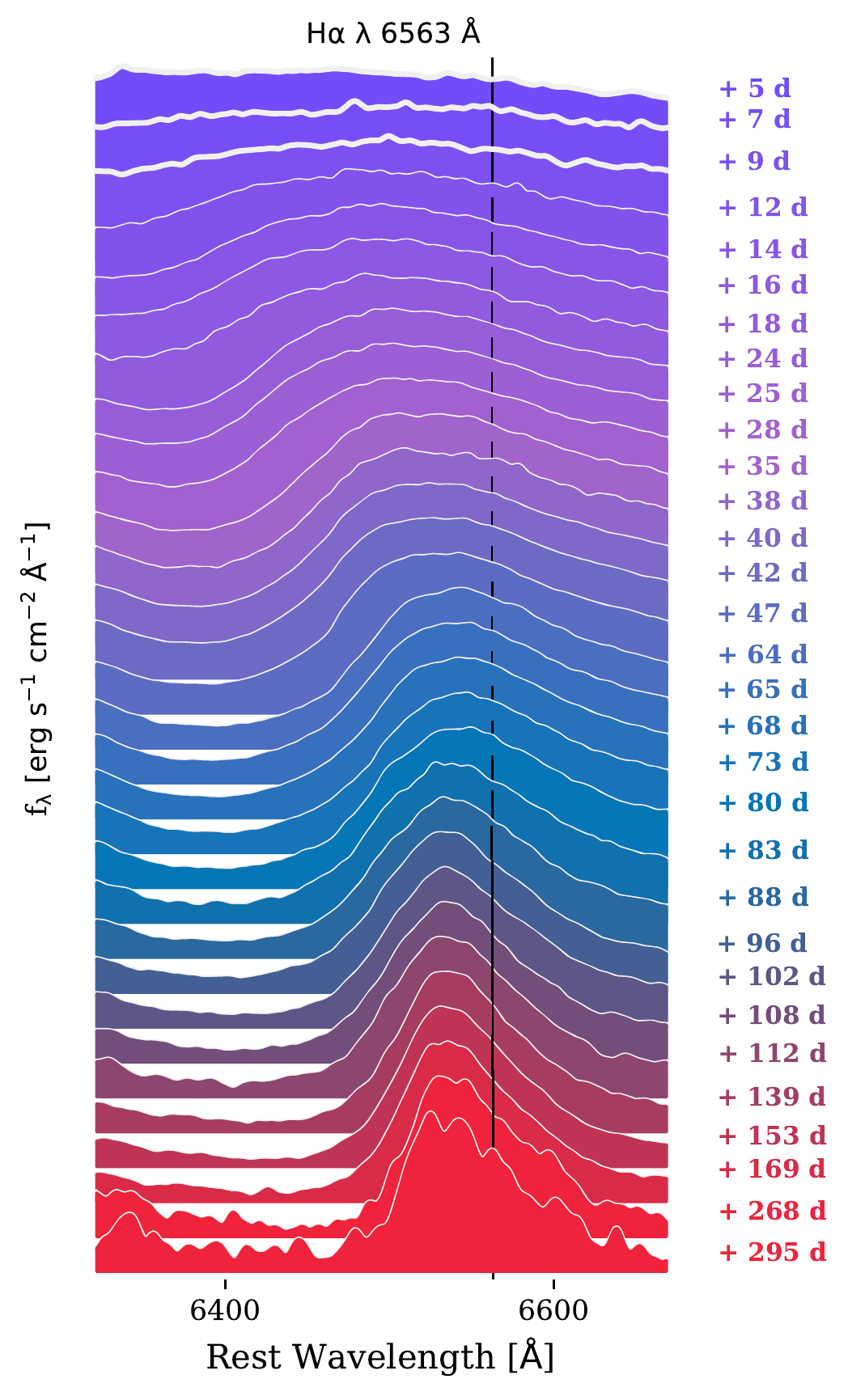}}
    \caption{Focused view of spectral time series for H$\alpha$ line in SN~2018gj. The black line marked is the rest wavelength for H$\alpha$. The spectra are corrected for host redshift, and it is evident that the peak of emission features never reaches the rest wavelength. }
    \label{fig:spectraEvolutionH2}
\end{figure}

\begin{figure}
	 \resizebox{\hsize}{!}{\includegraphics{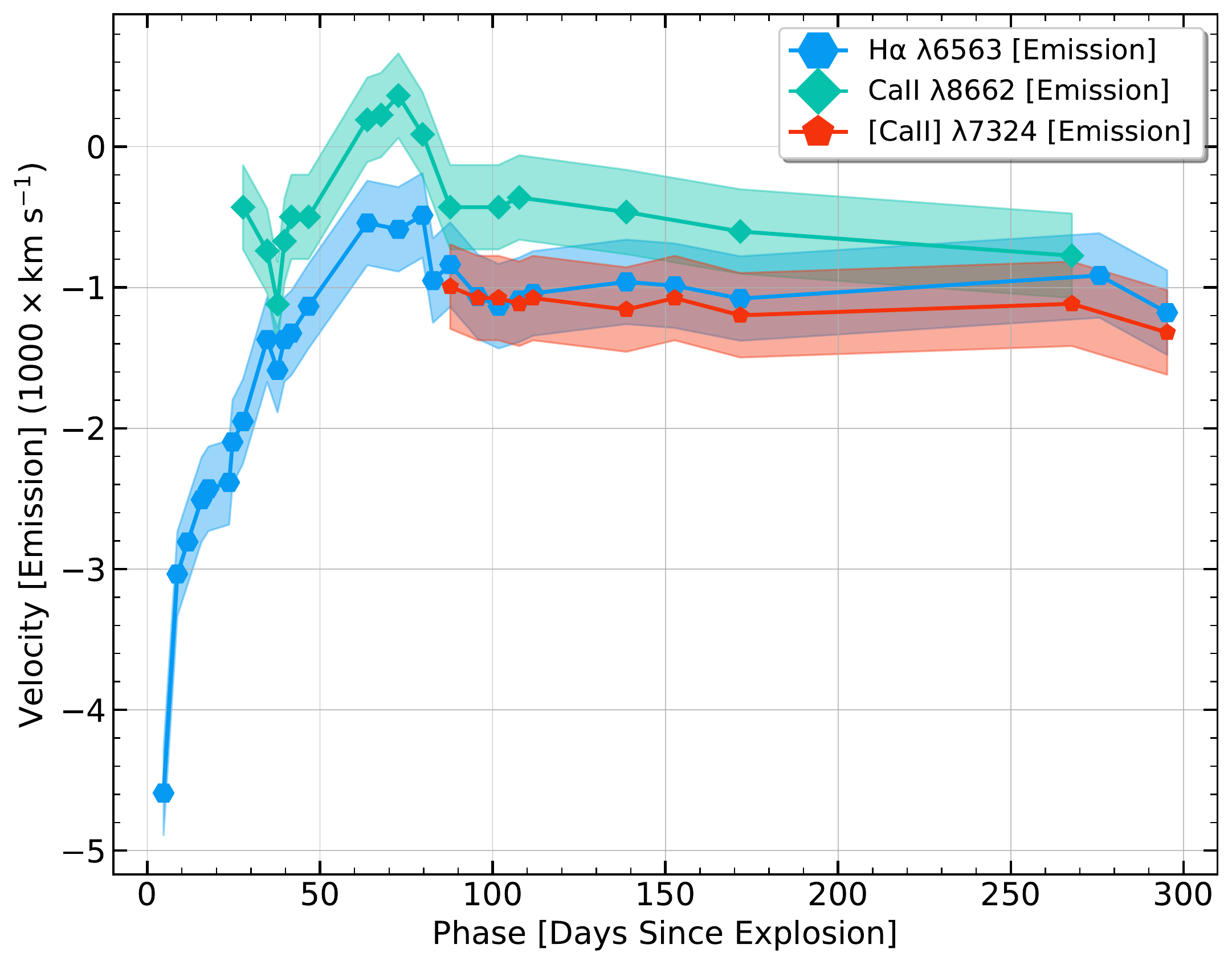}}
    \caption{Velocity evolution of blue-shifted $\rm H\alpha$ emission peak. Velocity evolution for emission features obtained using other ions has also been plotted.}
    \label{fig:H_alpha_emission}
\end{figure}

We observed blueshift in the emission peaks in the spectral evolution of SN~2018gj. In Figure~\ref{fig:spectraEvolutionH2}, the region around H$\alpha$ has been plotted, showcasing this persistent blue shift in the H$\alpha$ emission peak. The H$\alpha$ emission peak is shifted by $\rm \sim\,4500~km~s^{-1}$ around +10~d, which decreases monotonically till the end of the plateau around +75~d where it reaches a value $\rm \sim\,500~km~s^{-1}$ but never reaches rest wavelengths (Figure~\ref{fig:H_alpha_emission}). Instead, we observed the shift to increase during the transition phase and settle on a value of $\rm - 1000~km~s^{-1}$. The blue shift is seen till the last available spectrum ($+$295 d). The shift is not only observed in the H$\alpha$ but is also seen in other lines with similar values.


\subsection{Progenitor Mass Estimates Using Nebular Lines}

\begin{figure}
	 \resizebox{\hsize}{!}{\includegraphics{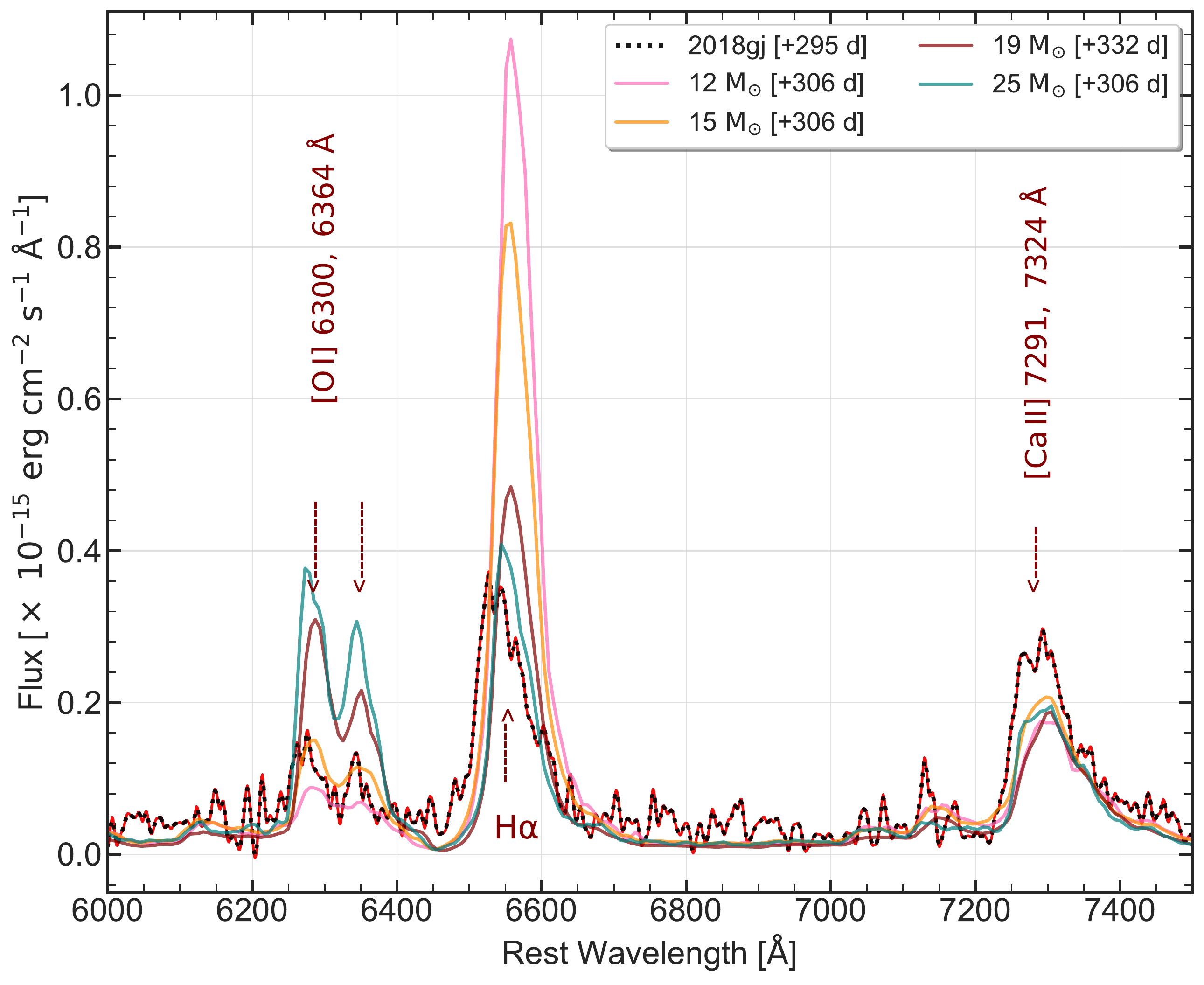}}
    \caption{Late phase (+295~d) spectrum of SN~2018gj compared with the model spectra around similar phase to estimate the initial mass of the progenitor. The model spectra are obtained from \citet{Jerkstrand2017}.}
    \label{fig:Jerkstrand}
\end{figure}

In addition to light curve modelling, there are other independent methods for determining the progenitor's mass. One such approach is to utilize the nebular phase spectra, which can provide insight into the metallic lines that arise from stellar nucleosynthesis. \citet{2014jerkstrand} demonstrated that the late-phase lines of [\ion{O}{1}]\,$~6300,\, 6364\,\AA$ and [\ion{Ca}{2}]\,$~7291,\, 7324\,\AA$ can serve as proxies for progenitor mass. To constrain the progenitor mass of SN~2018gj, we compared the nebular phase spectrum at +295~d with model spectra from \citet{Jerkstrand2017}. The model spectra for progenitor masses of 12, 15, 19, and 25\,$\rm M_\odot$ have been scaled for the $\rm ^{56}Ni$ mass and the distance of SN~2018gj (in contrast to 5.5 Mpc for the model spectra) (Figure~\ref{fig:Jerkstrand}). To account for the difference in phase between the model spectra and the observed spectrum, the observed spectrum was scaled by the brightness difference due to dissimilarity in phases determined from the characteristic time scale ($t_c$) obtained from the $\rm ^{56}Ni$-decay powered phase of the light curve using $f_{corr}=f_{obs}/(1-e^{-(t_c/phase)})$ \citep{2019ApJ...882...68Singh}. 

The comparison of [\ion{O}{1}]\,6300\,\AA, 6364\,\AA\ line fluxes of the observed spectra with the spectral models suggests a progenitor of mass $\rm \leq$\,15\,$\rm M_{\odot}$. However, the observed $\rm H\alpha$ flux is relatively weak as compared to the 15\,$\rm M_{\odot}$ progenitor, indicating a partially stripped hydrogen envelope in SN~2018gj. In core-collapse SNe, the mass of calcium synthesized in the explosion is insensitive to the progenitor's zero-age main-sequence mass, whereas the mass of synthesized oxygen depends on it. Hence, the  [\ion{Ca}{2}]\,/\,[\ion{O}{1}] flux ratio is a useful indicator of the progenitor mass \citep{1989fransson}. The lower the value of the ratio, the heavier the progenitor. As seen in the model spectra presented in Figure~\ref{fig:Jerkstrand}, the [\ion{Ca}{2}]$~7291,\, 7324\,\AA$ line from different mass models have similar line strength whereas it is differentiable in the case of [\ion{O}{1}] $~6300,\, 6364\,\AA$ and increases with the progenitor masses. In the case of SN~2018gj, the [\ion{Ca}{2}] lines are stronger than the model spectra, whereas [\ion{O}{1}] lines are much weaker. Therefore the [\ion{Ca}{2}]\,/\,[\ion{O}{1}] line flux ratio is much larger than one indicating a low mass progenitor.

\section{Hydrodynamical Modelling}
\label{sec:Progenitor}

\begin{table*}
\centering
\caption{Parameters for pre-SN/Explosion progenitor models evolved from \texttt{MESA} that were used to generate model light curves in \texttt{STELLA}} 
    \begin{tabular}{c|c|c|c|c|c|c|c|c|c|c|c}
    \hline
    \hline
      \multicolumn{10}{|c|}{\textbf{Pre-Supernova Parameters}} & \multicolumn{2}{|c|}{\textbf{Explosion Parameters}}\\

      \hline
      \hline
      $ M_i$ &$ \alpha_{Dutch}$ & $\rm M_f$ &$\rm M_{H-rich} $ & $\rm M_{He-core}$ &$\rm M_{Fe-core}$&$\rm logT_{eff}$&$\rm logL$ & Age & Radius & $ E_{Exp}$ & $ M_{Ni}$ \\
      ($\rm M_\odot$) & &($\rm M_\odot$) &($\rm M_\odot$) &($\rm M_\odot$)  &($\rm M_\odot$) & $\rm T_{eff}(K)$ & $\rm L(erg~s^{-1})$ & (Myr) & ($\rm R_\odot$) & $\rm (10^{51}~erg) $&($\rm M_\odot$) \\
      \hline
      \multirow{5}{1em}{13} & 3.0&9.1 & 5.35 & 3.75  & 1.50 & 3.57 & 4.83 & 16.7 & 620 &0.4&0.025 \\
      & 4.0&8.2 & 4.54 &3.63 & 1.53 &3.57  & 4.80 & 16.7  & 609 & 0.4 & 0.025\\
      & 4.5&7.3 & 3.66 & 3.65  &1.48  &3.51  &4.77  &16.7 &773 & 0.4 & 0.030\\
      & 5.0&6.9 & 3.26 & 3.60  &1.52  &3.51  & 4.80 &16.8 &794 & 0.2 & 0.027\\
      & 5.0&6.9 & 3.26 & 3.60  &1.52  &3.51  & 4.80 &16.8 &794 & 0.3 & 0.027\\
      & \bf{5.0} & \bf{6.9} & \bf{3.26} & \bf{3.60}  &\bf{1.52}  & \bf{3.51}  & \bf{4.80} &\bf{16.8} &\bf{794} & \bf{0.4} & \bf{0.028}\\
      & 5.5&6.3 & 2.69 & 3.57 &1.61  &3.52  &4.72  &16.9 &705 & 0.3 & 0.025\\
      \hline
      \hline
    \end{tabular}
    \label{tab:MESA_preSN}
\end{table*}

\begin{figure}
	 \resizebox{\hsize}{!}{\includegraphics{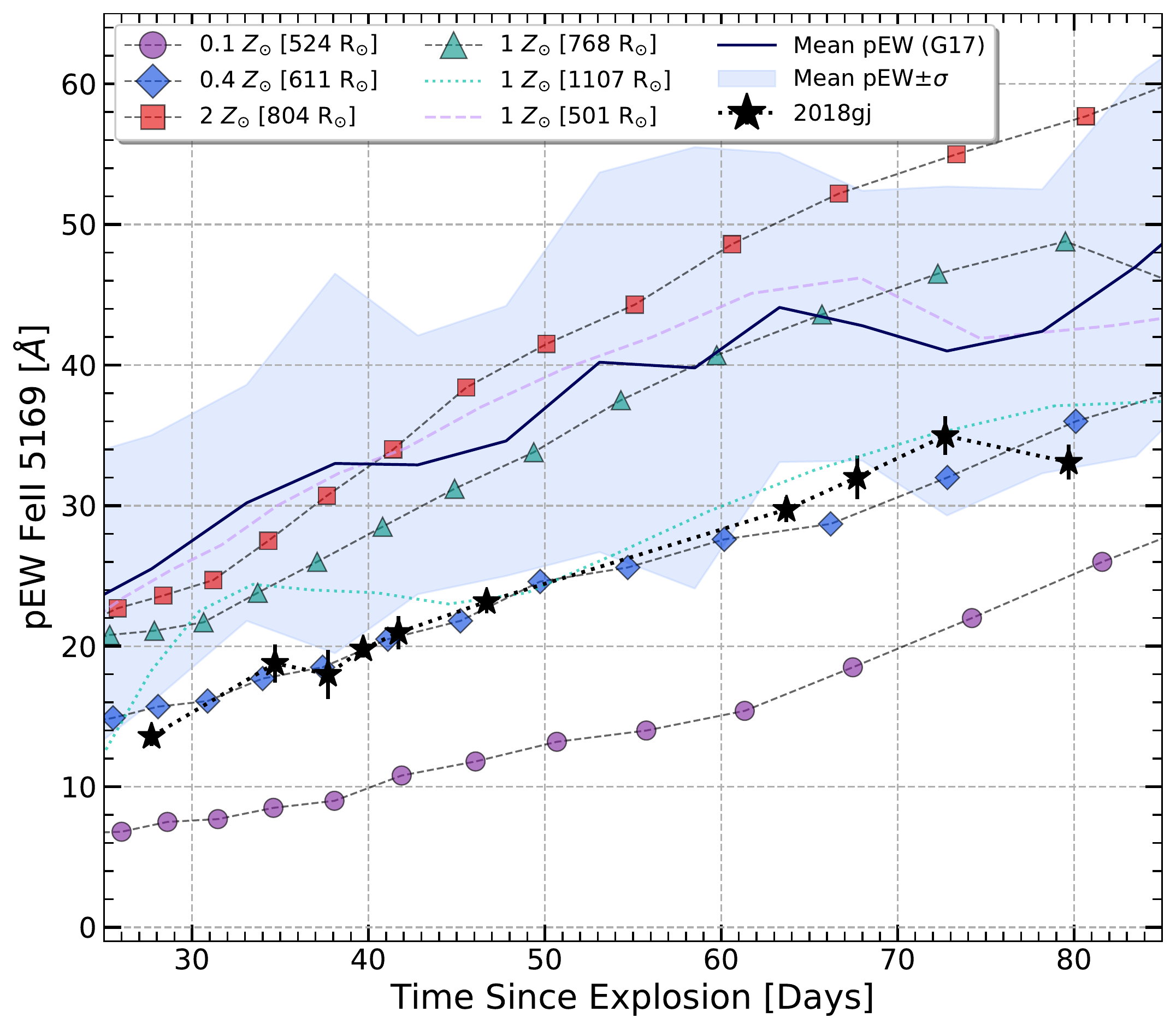}}
    \caption{Temporal evolution of pseudo-equivalent width (pEW) for \ion{Fe}{2} $\rm~5169~\AA$ line obtained using optical spectra. Other markers represent the pEWs obtained from models given in \citet{2013MNRAS.433.1745DpEW}. The solid blue line and the shaded region around it represent the mean pEW and the corresponding 1-$\sigma$ scatter about the mean for a larger sample of Type II SNe given in \citet{Gutierrez2017_TypeIISample}.}
    \label{fig:pEWMean}
\end{figure}

\begin{figure}
	 \resizebox{\hsize}{!}{\includegraphics{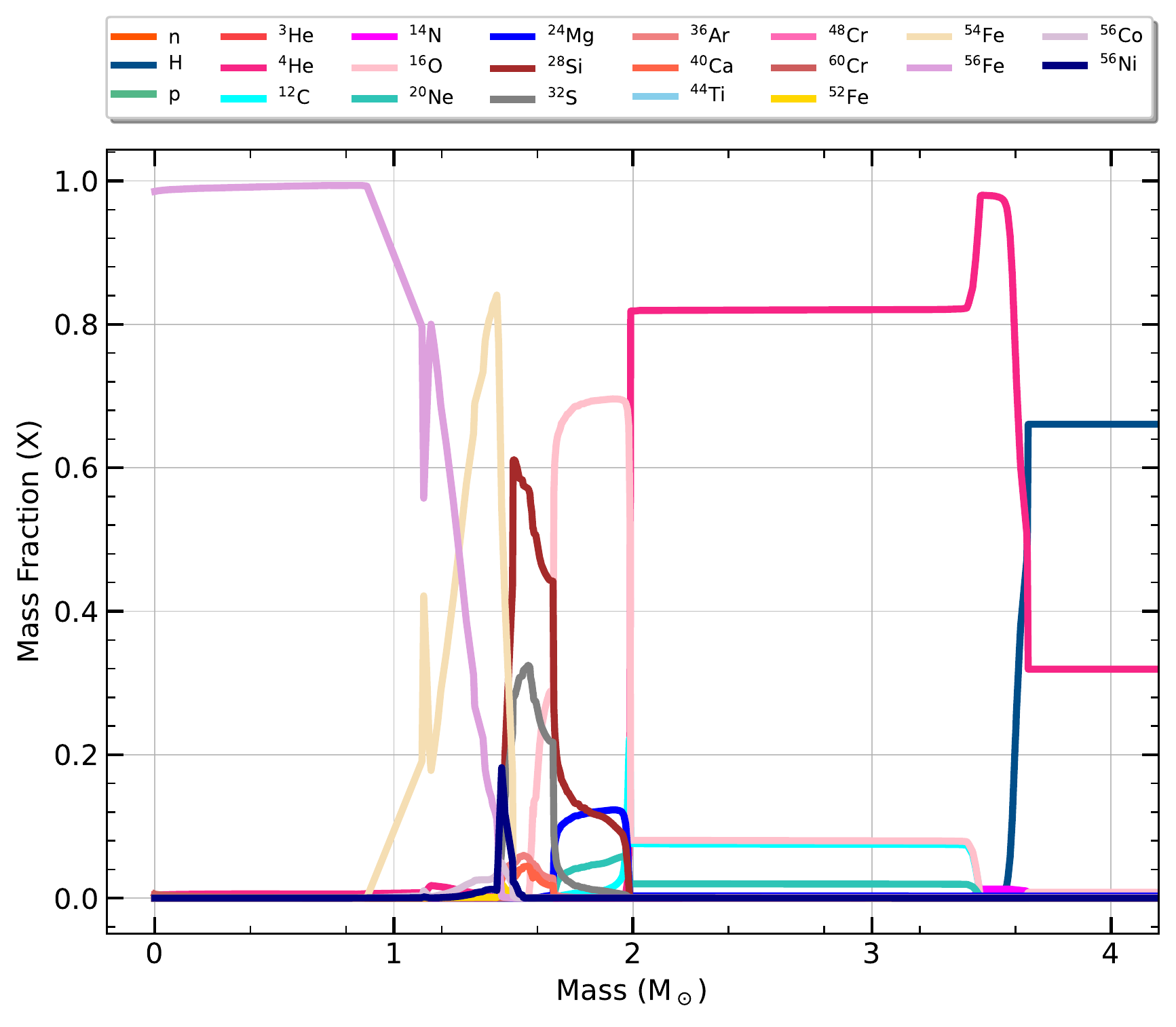}}
  \resizebox{\hsize}{!}{\includegraphics{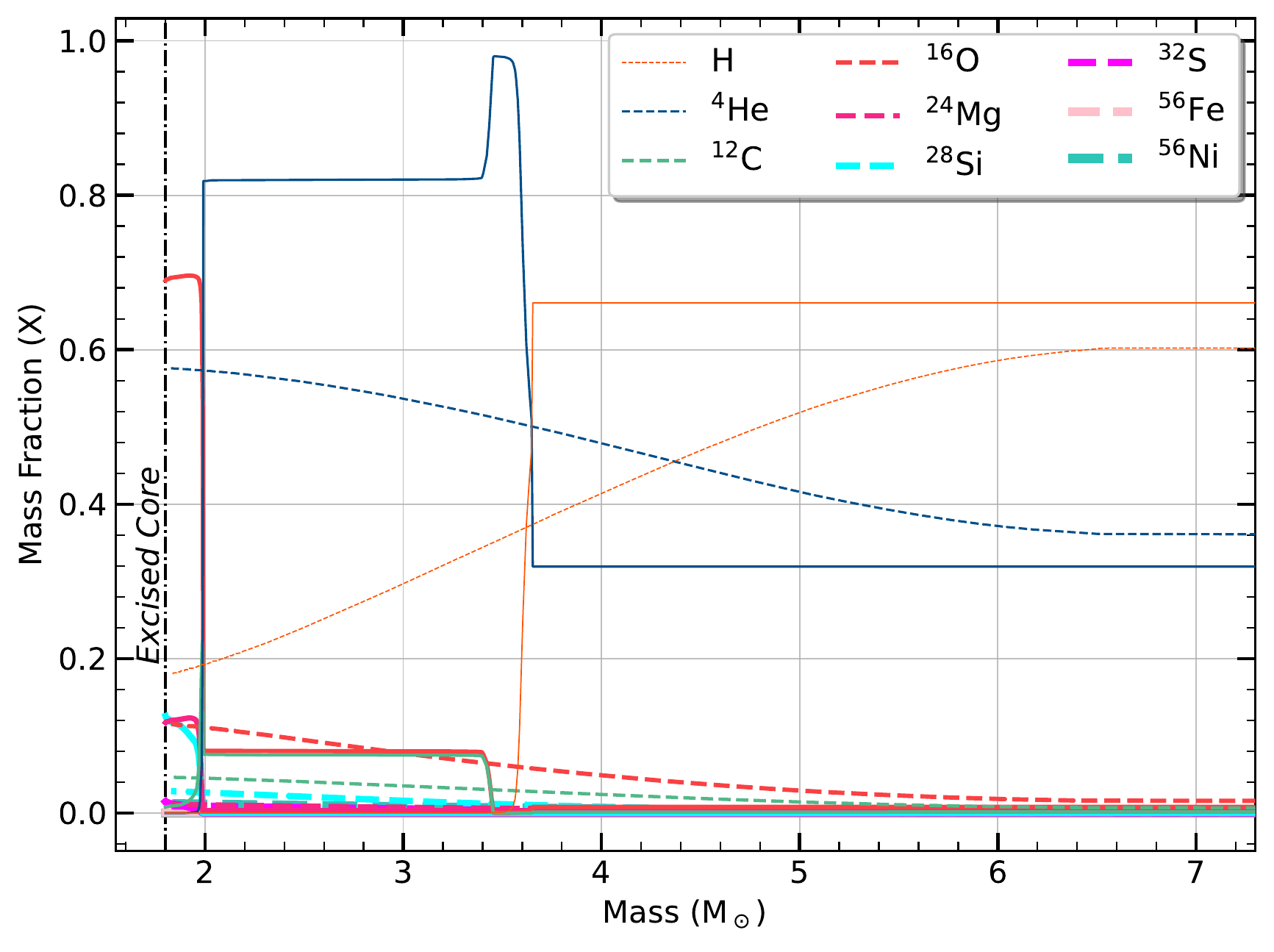}}
    \caption{{\bf Top:} One dimensional representation of mass fractions for 13~$\rm M_\odot$ ZAMS model with final mass 7.3~$\rm M_\odot$. The elements are part of the nuclear reaction rates network used in the model's evolution. The abscissa is shown till 4~$\rm M_\odot$ as the trend followed beyond it is the same for the outer hydrogen envelope. Bottom: Effect of mixing with the implementation of Duffell RTI on the ejecta structure just before the breakout is shown for some of the prominent elements for the same model as in the Top panel. }
    \label{fig:mesa_structure}
\end{figure}

The lower bound for progenitor mass obtained using the semi-analytical light curve modelling and the independent progenitor mass estimated using the nebular spectrum comparison with model spectra suggest  that SN~2018gj resulted from a low-mass RSG progenitor with ZAMS mass ranging from 10-15~$\rm M_\odot$. Model light curves of normal Type IIP SNe have been extensively studied \citep{2010MNRAS.408..827D, 2016Sukhbold, 2018PASA...35...49E}, and in some cases, refined analytical equations are provided to get estimates on the progenitor properties \citep{2019ApJ...879....3GGOLD}. We compare the observables obtained for SN~2018gj with the analytical equations \citep{2019ApJ...879....3GGOLD} obtained for Type IIP and find that for a specific radius range of RSG ($\rm 300-1200~R_\odot$), the equations hint at a shallow ejecta mass  ($\rm ~0.5-2.5~M_\odot$) and explosion energy ($\rm 0.4-0.03~foe$) (see Appendix~\ref{appendix:scaling}). These ejecta masses are in disagreement with our previous estimates obtained using semi-analytical modelling and nebular spectra comparisons. Even if we use radii obtained from the semi-analytical modelling in the scaling relation, the values obtained for the ejecta mass and explosion energy are not similar to those obtained from the semi-analytical modelling. But as we go for much smaller radii ($\sim$200~R$\odot$), the ejecta mass (4.5~M$\odot$) and explosion energy ($~\rm \sim 1~foe$) increase reaching closer to the values obtained using semi-analytical modeling. However, it is noted here that the analytical equations are calibrated for the SNe IIP that show a normal plateau of $\sim100$~days and may not necessarily be valid for short plateaus as also indicated by \citet{2021Hiramatsu}.

\begin{figure*}
	 \resizebox{\hsize}{!}{\includegraphics{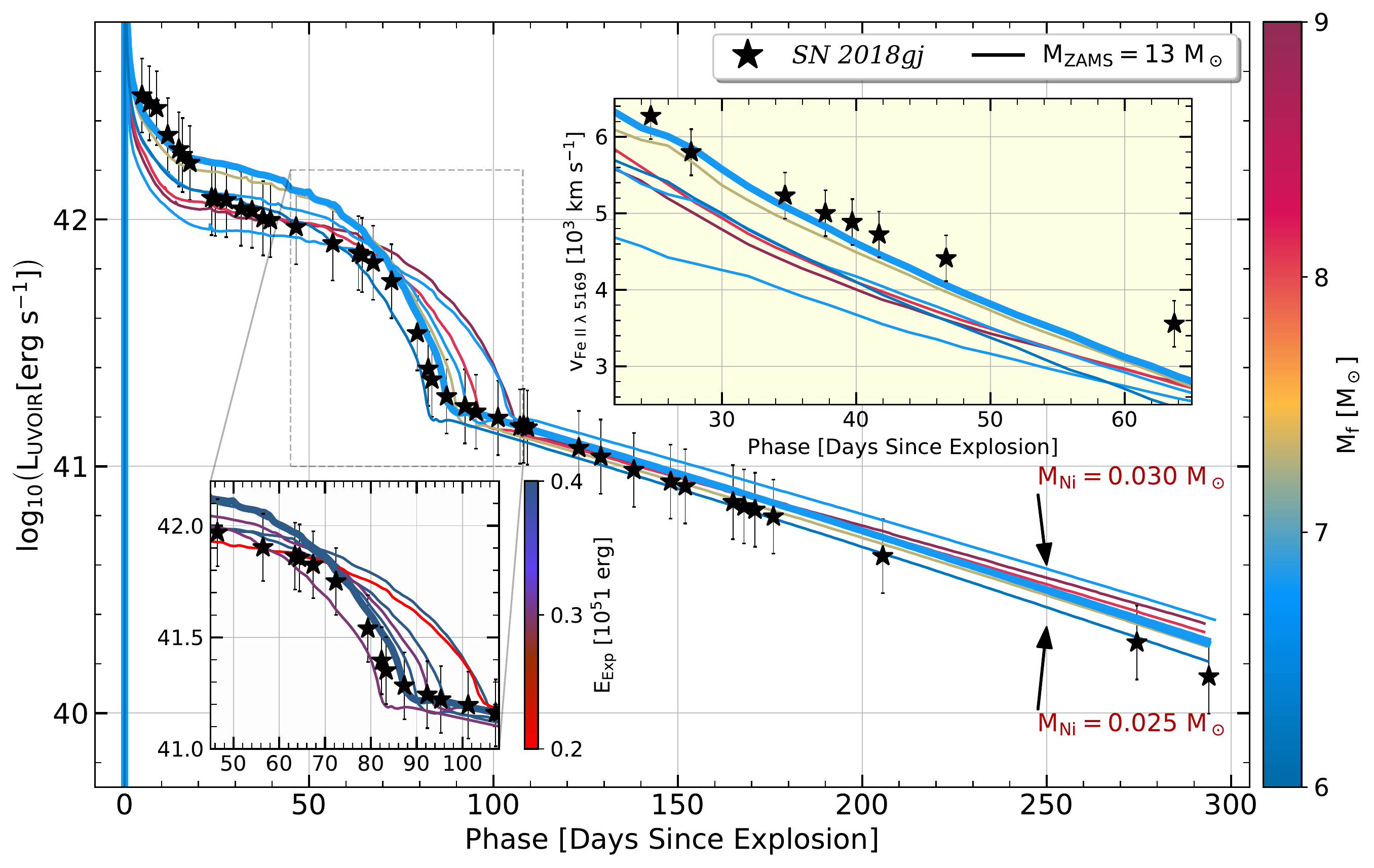}}
    \caption{ Variations in 13 $\rm M_\odot$ ZAMS model using different parameters to achieve a shorter plateau length. Zoomed out a plot in the bottom left shows the variation in explosion energy for different model light curves around the plateau transition. The second plot in the right inset shows the corresponding Fe 5169 velocities obtained using models. The thicker line represents the model where the expansion velocity could be matched with the observed velocities.}
    \label{fig:mesa_stella}
\end{figure*}

A more robust way forward is to perform a complete hydrodynamical modelling to better understand the progenitor, its evolution history, and other SN explosion parameters. We perform the modelling using the Modules for Experiments in Stellar Astrophysics \citep[\texttt{MESA} version r-15140,][]{Paxton2011, Paxton2013, Paxton2015, Paxton2018, Paxton2019} and \texttt{STELLA} \citep{Blinnikov2004, Baklanov2005, Blinnikov2006} packaged within \texttt{MESA}. It was compiled using MESA SDK \citep[version x86\_64-linux-20.12.1,][]{richard_townsend_2020_4587206}. The modelling setup follows the values prescribed in \citet{Farmer}. In lower mass models, we had increased the $max\_model\_number$ in each $inlists$ to accommodate longer evolution times. We fixed the overshooting values to default settings ($f=0.01, f_0=0.005$). We set the $varcontrol\_target=10^{-4}$ for the convergence of models with higher mass loss. Some other basic setup parameters are as follows: the nuclear reaction rates network used `approx21\_cr60\_plus\_co56.net', provided within \texttt{MESA}. These rates are primarily from the Nuclear Astrophysics Compilation of Reaction rates, \citep[NACRE,][]{1999nacre} and the Joint Institute for Nuclear Astrophysics, JINA reaction rates \citep[REACLIB,][]{2010JINA}. The mixing length parameter (\texttt{MLT\_option}) defaults to $Henyey$ \cite{henyey1965}, with $\rm \alpha_{MLT} = 1.5-2.0$, where $\rm \alpha_{MLT}$ is the ratio of mixing length to the pressure scale height ($\rm = P/g\rho$). Other than the models with 19~$\rm M_\odot$ and 13~$\rm M_\odot$ ($\rm \alpha_{dutch}=3.0$) where $\rm \alpha_{MLT}$ is set to default value of 1.5, we set it to 2.0 for the remaining models. Cool and hot wind schemes {\bf($\alpha_{dutch}$)} for the Red Giant Branch or Asymptotic Giant Branch phase are considered `Dutch', combining works by many Dutch authors. The primary combination chosen is based on the work by \citet{dutch1}. Typically, if the surface \ion{H}{0} has a mass fraction $<$ 0.4 and a $\rm T_{eff}$ $>$ $\rm 10^4\,K$, the scheme used is from \cite{vink1}, otherwise, it is from \cite{nugis1}. The default $Ledoux$ criterion is used to determine the position of the convective boundaries. 

The light curve modelling in MESA+STELLA was achieved in the broad steps: progenitor evolution, synthetic explosion, shock propagation, shock breakout and ejecta evolution. In the first step, we used \texttt{"make\_pre\_ccsnIIp"} example provided in \textit{test suites} to evolve a pre-main sequence star until the development of the iron core further leading to its rapid infall. Given that there is no solution to achieve a spontaneous explosion in MESA, the \textit{test suite}, namely \texttt{"ccsnIIp"}, provides an alternative methodology to achieve the desired explosion. Primarily, a thin layer (mass $\rm \sim 0.01~M_\odot$) at the inner boundary is injected with a tremendous amount of energy over a brief period (5~ms) to achieve the synthetic explosion. Finally, the shock is propagated with various layers, and the structure and hydrodynamical parameters are obtained just before the shock breakout as input to the \texttt{STELLA} \citep{Paxton2018}. Further details of the steps adopted and the parameters adjusted from evolution to explosion and handover to \texttt{STELLA} could be obtained from \citep{Farmer, 2022ApJ...930...34T}. We used 400 zones for \texttt{STELLA} with 40 extra zones in case of CSM. For the case of bolometric light curves, we used 40 frequency bins. However, in the case of UBVRI light curves, we had estimated 13$\rm M_\odot$ cases for 120 frequency bins for better resolution. Another parameter crucial in modelling is the metallicity which could affect wind-driven mass losses, the mass of the hydrogen envelope, and the line-profile signatures in Type II SNe \citep{2013MNRAS.433.1745DpEW}. Metallicity becomes more significant in the case of short plateaus due to extensive wind mass losses and smaller hydrogen envelopes. Since SN~2018gj is far from its host galaxy centre, we tried to image the region around the SN in search of possible nearby \ion{H}{2} using a narrow band H$\alpha$ filter. Unfortunately, we could not detect any such region for sufficiently long exposures. To crudely estimate metallicity, we utilized the pseudo-equivalent width (pEW) evolution of \ion{Fe}{2}$\rm~\lambda~5169~\AA$. We compared it with the models presented in the \citet{2013MNRAS.433.1745DpEW}. Figure~\ref{fig:pEWMean} shows the time evolution of pEW  of \ion{Fe}{2}$\rm~\lambda~5169~\AA$ along with the models presented in \citeauthor{2013MNRAS.433.1745DpEW}. It also shows the mean and 1$-\sigma$ scatter in corresponding values for a larger Type II SNe sample \citep{Gutierrez2017_TypeIISample}. From these models, we found that two models with 0.5$\rm ~Z_\odot$ and $\rm Z_\odot$ matched with the pEW obtained in the case of SN~2018gj. Hence, we fix the metallicity of all the models to be of solar values. 


\begin{figure}
	 \resizebox{\hsize}{!}{\includegraphics{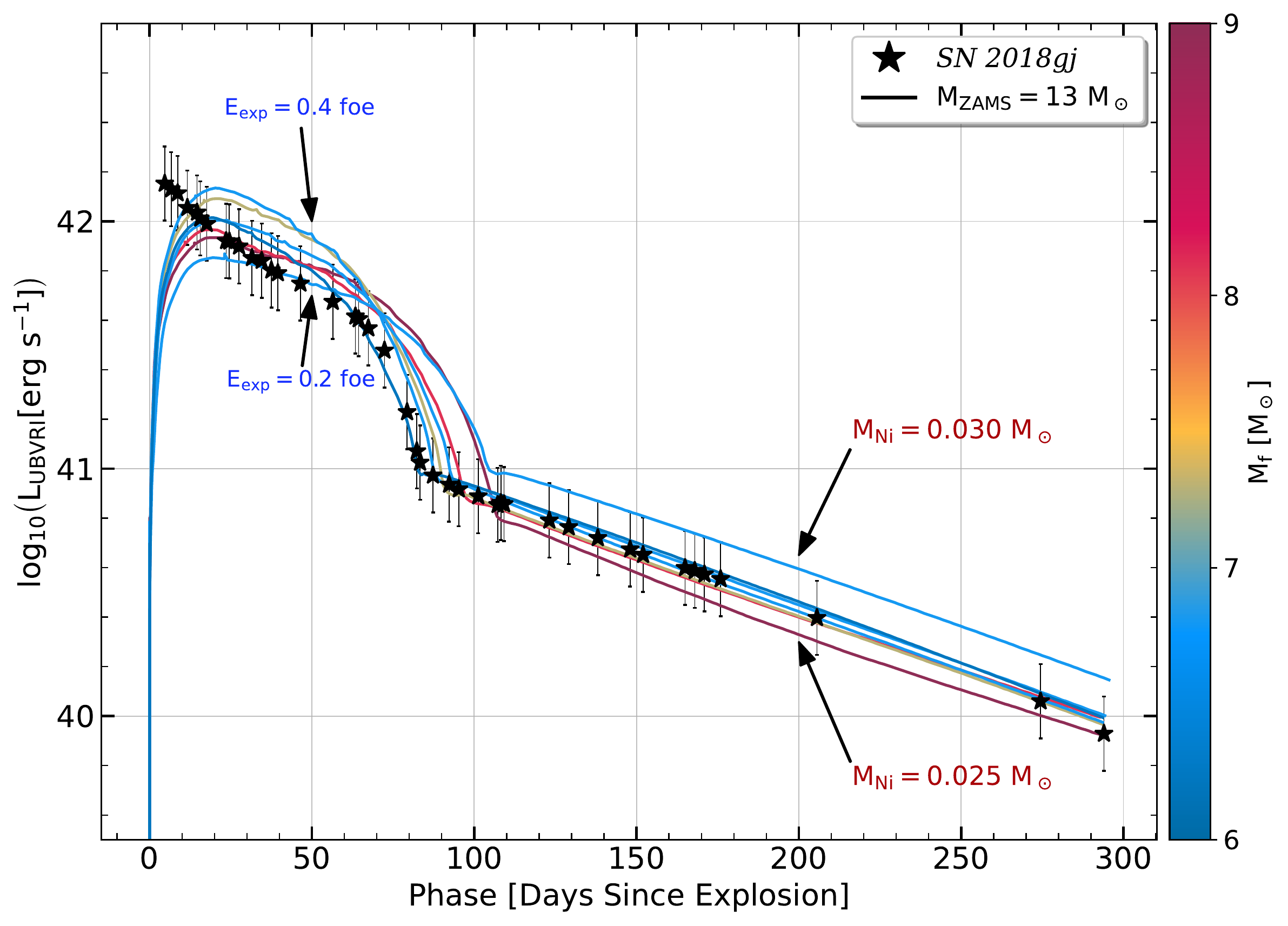}}
    \caption{Optical bolometric luminosities obtained using \texttt{MESA+STELLA} are plotted along with the optical bolometric light curve of SN~2018gj. The initial rise is not fitting well in the optical regime.}
    \label{fig:mesa_stella_qbol}
\end{figure}

We tried to evolve progenitors with zero-age-main-sequence (ZAMS) masses 13 $\rm M_\odot$ and 14 $\rm M_\odot$ and extracted their pseudo-bolometric/bolometric light curve evolution after they explode. We checked for higher mass models also (see Appendix~\ref{appendix:19M})  We found out that with standard mass loss by winds, none of the models was able to reproduce a short plateau. In most cases, the plateau duration was typical of normal Type IIP SNe. However, we could get shorter plateaus as we enhanced the mass loss rate through winds using the \textit{ wind scaling factor, $\alpha_{dutch}=3.0-5.5$}. As the plateau duration primarily depends on the hydrogen envelope mass, achieving shorter plateaus from each of these progenitor masses with the correct mass loss was possible. However, the simultaneous match to the expansion velocities was not achieved in the models for 14 $\rm M_\odot$ progenitors. In the case of 13 $\rm M_\odot$ models, we were able to match the light curve and expansion velocities up to the initial 50 days. Figure~\ref{fig:mesa_structure} gives the final composition for one of the 13~$\rm M_\odot$ models representing the elements used in the progenitor structure. It also shows the mixing effect on the ejecta composition with the implementation of Duffell RTI in MESA \citep{2016ApJ...821...76D}. The mass fractions beyond  4~$\rm M_\odot$ are very similar with no recognizable changes and, therefore, are not shown in the top panel of Figure~\ref{fig:mesa_structure}. Although elements are mixed at different mass coordinates, the core and outer structure fairly consist of iron and hydrogen, respectively. From the current understanding of single star evolution, high mass RSG ($\rm >20~M_\odot$), with enough mass loss, could give a smaller plateau as obtained in the works by \citet{2010MNRAS.408..827D, 2021Hiramatsu}. To explore the possibility of a high mass progenitor, we also attempted to generate models using 19~$\rm M_\odot$ progenitors, as this is the upper limit for directly detected progenitors.
It is possible to obtain smaller plateau lengths with lower mass loss rates, but these models failed to reproduce the velocity evolution of the ejecta [Appendix~\ref{appendix:19M}]. Properties of some of the pre-SN~progenitors are provided in Table~\ref{tab:MESA_preSN}, giving details of the initial and final masses of the progenitors.


Further, masses of helium and iron core present during evolution are also mentioned in the table. In addition to the used parameters, Table~\ref{tab:MESA_preSN} also lists the various properties of the pre-SN star, viz. effective temperature, luminosity, age, and radius. We only show those models where we could achieve smaller plateau lengths. Detailed descriptions and analyses of these individual models are beyond the scope of this work. As expected from the initial mass of the progenitor models, there are not many differences in the pre-SN structure apart from the mass difference of the hydrogen envelope. All the models with the same initial mass have similar evolution times, effective temperatures, core masses, and luminosities.

We attempted to generate the model light curves to match the observed UVOIR bolometric luminosity. After achieving a desired plateau length in the model light curves, the mass of  synthesized nickel and explosion energy were constrained by varying the nickel mass ($x\_ctrl(12)$) and explosion energy ($inject\_until\_reach\_model\_with\_total\_energy$) parameters during the explosion and shock propagation. Figure~\ref{fig:mesa_stella} shows all the models for 13 $\rm M_\odot$ with plateau lengths $\rm 80\pm10~d$. The bolometric luminosity of SN~2018gj is over-plotted. In the model, the explosion energy and nickel mass are well constrained, however, the initial peak (s1) is slightly under-luminous. Instead of comparing the bolometric luminosity, we compare the observed  pseudo (optical) bolometric luminosity with modelled pseudo-bolometric luminosity ($\rm L_{UBVRI}$), which reveals the under-luminous s1 phase more prominently. We discuss the possible presence of CSM interaction in the subsequent section. The nickel mass obtained with the hydrodynamical modelling corroborates the earlier mass estimates through various techniques. The explosion energy obtained is slightly less than the semi-analytical modelling estimates.

\section{Discussion}
\label{sec:Discussion}

\subsection{Blueshifted Emission}

As described earlier in Section~\ref{subsec:blushift}, we observed the emission peaks were blueshifted, and these shifts were observed till the late phase. The blue-shifted emission during the photospheric phase has been observed and discussed explicitly in many works \citep{ 2011ApJ...731...47ABS2007it, Bose2015_2013ab, 2018MNRAS.479.2421DBS2015ba}. It has been established in \citet{2014MNRAS.441..671ABSAnderson} that this feature is typical to Type IIP SNe during the early/photospheric phase. It was concluded by \citet{1988SvAL...14..334CBS1} that these blue-shifted emission peaks during the photospheric phase are due to the diffuse reflection of the photosphere's resonance radiation. Primarily in all the cases, the shifts are only present up to the late photospheric phase or early nebular phase except in the case of SN~2007it \citep{2011ApJ...731...47ABS2007it} where it has been observed till 150~d post-explosion. \citet{2014MNRAS.441..671ABSAnderson} utilizing the Type IIP modelled spectra from \citet{2013MNRAS.433.1745DpEW} showed the shifts in emission peaks vanish after photospheric phase and the emission peaks are observed at rest wavelengths. Interestingly, the shifted peak is reported for H$\alpha$ and not any other lines. 

The blue shift in the emission peaks observed in the spectral evolution of SN~2018gj is a typical feature for Type IIP SNe during the photospheric phase. But the intriguing aspect is the presence of these shifts until the late nebular phase and are not just limited to the prominent H$\alpha$ feature. Observance of such shifts during the late nebular phase is rare in most of the usual Type IIP SNe studied in the literature. The blueward shift might not be physical but apparent and can be explained from the argument presented in \citet{2014MNRAS.441..671ABSAnderson} regarding ejecta geometry and its composition. One of the critical factors is the changes in the opacity values. Opacity within the photosphere and above depends on various physical processes, viz. density, composition, and  degree of ionization within the ejecta \citep{2017hsn..book..769S}. During the photospheric phase, the density structure in Type IIP ejecta is much steeper. It enhances the confinement of the emission/absorption line. Further, it aggravates the concealment of the receding portion of the ejecta, hence biasing the blueward line emission for a distant observer during photospheric phase\citep{2014MNRAS.441..671ABSAnderson}.

But the above reasoning might not be valid during the nebular phase when the ejecta behaves like an emission line nebulae. Other radiative transfer effects might come into play, especially due to free electrons, photo-ionization, or the presence of dust \citep{2017hsn..book..795J}. During the nebular phase, the amount of electron scattering is relatively low, with an optical depth of $\tau_e \leq 1$. Most photons will not be scattered or will only be scattered once. As a result, the distortions in the line profile are not significant \citep{2017hsn..book..795J}. However, the scattering does cause a slight blue shift of the peak. For instance, when $\tau_e = 1$, the shift is approximately $\Delta \lambda/\lambda_0(V_{max}/c) =-0.13$, which corresponds to a velocity shift of $\rm 390~kms^{-1}$ for a line that is $\rm 3000kms^{-1}$ wide \cite{2017hsn..book..795J}. The blueshift observed in the spectra is much larger than the values obtained for typical opacity. Hence, this might not be the cause of the observed shifts.

The continuous absorptive opacity or the photon destruction (continuous absorption) by dust or photo-ionization could also cause a significant blue shift. For $\tau_e = 1$, the shift is approximately $\Delta \lambda/\lambda_0(V_{max}/c) =-0.31$, which corresponds to a velocity shift of $\rm \approx 900~kms^{-1}$ \citep{2017hsn..book..795J} which is significant and close to the observed values. But for these effects,  the presence of dust or enough optically thick material is required. As a considerable fraction of the hydrogen envelope is removed from the progenitor of SN~2018gj, the presence of optically thick material also does not seem plausible. However, the presence of pre-existing dust or early dust formation in the ejecta could be a possibility.  We do not find a convincing signature for the presence of dust in the ejecta. During the nebular phase, the light curve in the optical band is found to decline  faster than the $\rm ^{56}Co$ decay rate, which could be due to the light absorption by dust. In such a scenario, due to reprocessing of light by the dust particles, the light curve decline in the redder bands is expected to be slower. However, on the contrary, in SN~2018gj, light curves in the redder bands are found to decline much faster (Section~\ref{sec:light_curve_analysis}).

Since the supernova occurred  in the outskirts of the host  galaxy,  the intrinsically high velocity of the progenitor star towards the line of sight could also be the possibility. Although it is rare to find such high-mass hyper-velocity stars going rogue but could be possible, as observed in \citet{2015AJ....150..149E}.

\subsection{CSM Interaction?}

\begin{figure}
	 \resizebox{\hsize}{!}{\includegraphics{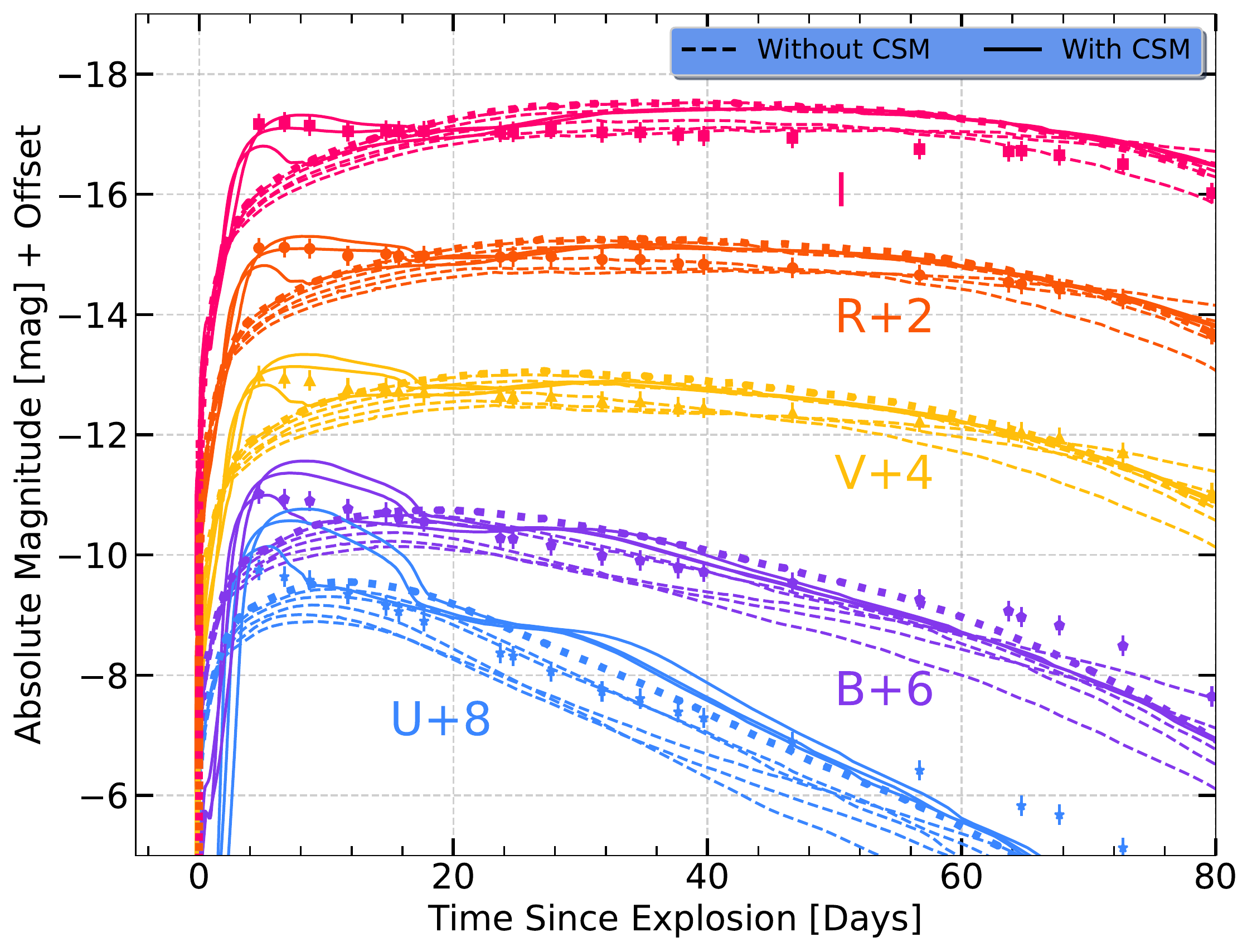}}
    \caption{The effect of adding CSM around the progenitor is prominently seen in the early stage and can explain the initial excess in the individual light curves in redder bands. The thicker dashed lines represent the light curves obtained using 120 frequency bins.}
    \label{fig:mesa_stella_csm}
\end{figure}

A piece of substantial evidence has been found in favour of early CSM interaction having a signature in the light curves. However, we do not see any interaction feature in the spectral evolution. When we try to fit pseudo bolometric luminosity with UBVRI bolometric luminosity from the models (see Figure~\ref{fig:mesa_stella_qbol}), we could see that the initial part does not fit that well until we introduce CSM interaction (see Figure~\ref{fig:mesa_stella_csm}).

In Figure~\ref{fig:mesa_stella_csm}, we introduce three CSM profiles with different wind evolution time and mass loss rates giving total masses 0.1, 0.15, 0.20 $\rm M_\odot$ with different extents. We observe that the initial light evolution could be explained with less than 0.15 $\rm M_\odot$ of CSM, which is close to the progenitor. The enhanced pre-SN wind was activated 10-20 years before the explosion.  

Since there is no evidence in spectra to corroborate the presence of CSM around the progenitor. Either the lack of CSM interaction evidence in spectra is due to some non-spherical geometry, or it might be the intrinsic feature of these SNe and warrant a further understanding of these light curve rise times.
\section{Conclusion}
\label{sec:Conclusion}

In this work, we presented a detailed investigation of a short plateau Type II supernova SN~2018gj, which exhibited a plateau lasting for $\rm \sim 70~d$ in its light curve. This plateau duration is significantly less than the characteristic plateau length of $\rm \sim 100~d$ for Type IIP SNe. We carried out detailed photometric observations in UV, optical, and NIR wavelengths and the detailed optical spectroscopic evolution from the photospheric phase to the nebular phase ($\sim$ 300~d from the explosion). The various light curve parameters were estimated, and the peak $V$-band magnitude was estimated as $\rm -17.0\pm0.1~mag$. Using the bolometric light curve mass of synthesized radioactive $\rm ^{56}Ni$ is estimated as {\bf$\rm 0.026\pm0.007~M_\odot$}. Spectroscopic comparison of SN~2018gj with other Type II SNe indicated it to be a normal Type II SN but with high H$\alpha$ velocities. Further, blueshift in the emission features during the late nebular phase is also reported. We carried out semi-analytical modelling, nebular phase spectral comparisons, and complete 1-D hydrodynamical modelling to ascertain ejecta mass, explosion energy, synthesized nickel mass, and details about the progenitor. The models favoured a low mass progenitor of ZAMS mass of $\rm < 13~M_\odot$, contrary to the higher mass RSG channels available in the literature. We found the mass of the hydrogen envelope to be only $\rm \sim 2.5-3.0~M_\odot$ and a total pre-SN mass $\rm \leq 7~M_\odot$.

\section*{Data Availability}
The majority of data are presented in tables. The spectra files are provided as data behind the figure. The inlists used for creating MESA models and associated models are available on Zenodo under an open-source Creative Commons Attribution license: \dataset[doi:10.5281/zenodo.7994631]{https://doi.org/10.5281/zenodo.7994631}.

Other data files, models, etc., will be shared with users upon reasonable request.

\begin{acknowledgments}
We are grateful to the anonymous referee for a thorough evaluation of the manuscript that helped in improving it. We thank the staff of IAO, Hanle, and CREST, Hosakote, that made these observations possible. The facilities at IAO and CREST are operated by the Indian Institute of Astrophysics, Bangalore. This research has made use of the High-Performance Computing (HPC) resources\footnote{\href{https://www.iiap.res.in/?q=facilities/computing/nova}{https://www.iiap.res.in/?q=facilities/computing/nova}} made available by the Computer Center of the Indian Institute of Astrophysics, Bangalore. This research made use of \textsc{RedPipe}\footnote{\url{https://github.com/sPaMFouR/RedPipe}} \citep{2021redpipe}, an assemblage of data reduction and analysis scripts written by AS. This work was supported by the Joint Research Project, Grant number JPJSBP120227709. AT acknowledges support from the Egyptian Science, Technology and Innovation Funding Authority (STDF) under the grant number 45779. DKS acknowledges the support provided by DST-JSPS under grant number DST/INT/JSPS/P 363/2022. This work was also partly supported by JSPS KAKENHI Grant Numbers JP50645512. This research has also made use of the NASA/IPAC Extragalactic Database (NED\footnote{\url{https://ned.ipac.caltech.edu}}), which is funded by the National Aeronautics and Space Administration and operated by the California Institute of Technology. This work has used data from the Asteroid Terrestrial-impact Last Alert System (ATLAS) project. The Asteroid Terrestrial-impact Last Alert System (ATLAS) project is primarily funded to search for near-earth asteroids through NASA grants NN12AR55G, 80NSSC18K0284, and 80NSSC18K1575; byproducts of the NEO search include images and catalogues from the survey area. This work was partially funded by Kepler/K2 grant J1944/80NSSC19K0112 and HST GO-15889, and STFC grants ST/T000198/1 and ST/S006109/1. The ATLAS science products have been made possible through the contributions of the University of Hawaii Institute for Astronomy, the Queen's University Belfast, the Space Telescope Science Institute, the South African Astronomical Observatory, and The Millennium Institute of Astrophysics (MAS), Chile.
\end{acknowledgments}

%

\vspace{5mm}
\facilities{IAO(HCT):2-m, Kanata(HONIR):1.5-m ATLAS, and Swift(UVOT)}


\software{Astropy \citep{astropy:2013, astropy:2018, astropy:2022} , emcee \citep{2013PASP..125..306F}, MESA \citep{Paxton2011, Paxton2013, Paxton2015, Paxton2018, Paxton2019}, STELLA \citep{Blinnikov2004, Baklanov2005, Blinnikov2006}, matplotlib \citep{Hunter:2007}, pandas \citep{mckinney-proc-scipy-2010, reback2020pandas}, numpy \citep{harris2020array}, scipy \citep{2020SciPy-NMeth}, Jupyter \citep{soton403913}, and seaborn \citep{Waskom2021}}


\newpage

\appendix

\section{DATA}
\label{appendix:data}
\begin{table*}[hbt!]
\centering
\caption{The $UBVRI$ photometric magnitudes of SN~2018gj.}
\label{tab:HCTphotometry}
\begin{tabular}{cccccccc} 
\hline
Date        &  JD      & Phase$^{\ast}$    &  $U$             & $B$              & $V$              & $R$        & $I$     \\     
(yyyy-mm-dd)& 2458000+ &(d)                &  (mag)           & (mag)            & (mag)            & (mag)      & (mag)            \\  
\hline   
2018-01-14 & 132.5 & $+$4.7   &  14.07 $\pm$ 0.01 & 14.75 $\pm$ 0.01 &  14.71 $\pm$ 0.01 &  14.55 $\pm$ 0.01 & 14.43 $\pm$ 0.01  \\    
2018-01-16 & 134.5 & $+$6.7   &  14.18 $\pm$ 0.01 & 14.84 $\pm$ 0.01 &  14.75 $\pm$ 0.01 &  14.54 $\pm$ 0.01 & 14.40 $\pm$ 0.01  \\
2018-01-18 & 136.5 & $+$8.7   &  14.24 $\pm$ 0.01 & 14.87 $\pm$ 0.01 &  14.79 $\pm$ 0.01 &  14.56 $\pm$ 0.01 & 14.46 $\pm$ 0.01  \\
2018-01-21 & 139.5 & $+$11.7  &  14.47 $\pm$ 0.01 & 15.00 $\pm$ 0.01 &  14.92 $\pm$ 0.01 &  14.68 $\pm$ 0.01 & 14.55 $\pm$ 0.01  \\
2018-01-24 & 142.5 & $+$14.7  &  14.66 $\pm$ 0.01 & 15.06 $\pm$ 0.01 &  14.92 $\pm$ 0.01 &  14.65 $\pm$ 0.01 & 14.54 $\pm$ 0.01  \\
2018-01-25 & 143.5 & $+$15.7  &  14.76 $\pm$ 0.01 & 15.14 $\pm$ 0.01 &  14.97 $\pm$ 0.01 &  14.68 $\pm$ 0.01 & 14.55 $\pm$ 0.01  \\
2018-01-27 & 145.5 & $+$17.7  &  14.92 $\pm$ 0.02 & 15.21 $\pm$ 0.01 &  14.99 $\pm$ 0.01 &  14.68 $\pm$ 0.01 & 14.55 $\pm$ 0.01  \\
2018-02-02 & 151.5 & $+$23.7  &  15.45 $\pm$ 0.02 & 15.49 $\pm$ 0.01 &  15.05 $\pm$ 0.01 &  14.70 $\pm$ 0.01 & 14.56 $\pm$ 0.01  \\
2018-02-03 & 152.5 & $+$24.7  &  15.50 $\pm$ 0.01 & 15.50 $\pm$ 0.01 &  15.05 $\pm$ 0.01 &  14.69 $\pm$ 0.01 & 14.56 $\pm$ 0.01  \\
2018-02-06 & 155.5 & $+$27.7  &  15.76 $\pm$ 0.02 & 15.61 $\pm$ 0.01 &  15.05 $\pm$ 0.01 &  14.69 $\pm$ 0.01 & 14.51 $\pm$ 0.01  \\
2018-02-10 & 159.5 & $+$31.7  &  16.09 $\pm$ 0.02 & 15.78 $\pm$ 0.01 &  15.14 $\pm$ 0.01 &  14.74 $\pm$ 0.01 & 14.57 $\pm$ 0.01  \\
2018-02-13 & 162.5 & $+$34.7  &  16.21 $\pm$ 0.02 & 15.86 $\pm$ 0.01 &  15.14 $\pm$ 0.01 &  14.74 $\pm$ 0.01 & 14.57 $\pm$ 0.01  \\
2018-02-16 & 165.5 & $+$37.7  &  16.43 $\pm$ 0.02 & 15.99 $\pm$ 0.01 &  15.23 $\pm$ 0.01 &  14.81 $\pm$ 0.01 & 14.62 $\pm$ 0.01  \\
2018-02-18 & 167.5 & $+$39.7  &  16.53 $\pm$ 0.02 & 16.05 $\pm$ 0.01 &  15.25 $\pm$ 0.01 &  14.82 $\pm$ 0.01 & 14.63 $\pm$ 0.01  \\
2018-02-25 & 174.5 & $+$46.7  &  16.93 $\pm$ 0.01 & 16.23 $\pm$ 0.01 &  15.33 $\pm$ 0.01 &  14.88 $\pm$ 0.01 & 14.66 $\pm$ 0.01  \\
2018-03-07 & 184.5 & $+$56.7  &  17.40 $\pm$ 0.01 & 16.51 $\pm$ 0.01 &  15.48 $\pm$ 0.01 &  15.00 $\pm$ 0.01 & 14.85 $\pm$ 0.01  \\
2018-03-14 & 191.5 & $+$63.7  &          --       & 16.70 $\pm$ 0.02 &  15.65 $\pm$ 0.01 &  15.12 $\pm$ 0.02 & 14.89 $\pm$ 0.01  \\
2018-03-15 & 192.5 & $+$64.7  &  17.99 $\pm$ 0.04 & 16.80 $\pm$ 0.01 &  15.65 $\pm$ 0.01 &  15.15 $\pm$ 0.02 & 14.88 $\pm$ 0.01  \\
2018-03-18 & 195.5 & $+$67.7  &  18.14 $\pm$ 0.03 & 16.94 $\pm$ 0.01 &  15.74 $\pm$ 0.01 &  15.23 $\pm$ 0.01 & 14.95 $\pm$ 0.01  \\
2018-03-23 & 200.5 & $+$72.7  &  18.69 $\pm$ 0.02 & 17.28 $\pm$ 0.01 &  15.99 $\pm$ 0.01 &  15.40 $\pm$ 0.01 & 15.10 $\pm$ 0.01  \\
2018-03-30 & 207.5 & $+$79.7  &          --       & 18.12 $\pm$ 0.02 &  16.66 $\pm$ 0.01 &  15.98 $\pm$ 0.01 & 15.58 $\pm$ 0.01  \\
2018-04-02 & 210.5 & $+$82.7  &          --       & 18.61 $\pm$ 0.02 &  17.11 $\pm$ 0.01 &  16.34 $\pm$ 0.01 & 15.93 $\pm$ 0.01  \\
2018-04-03 & 211.5 & $+$83.7  &          --       & 18.73 $\pm$ 0.02 &  17.20 $\pm$ 0.01 &  16.44 $\pm$ 0.02 & 16.10 $\pm$ 0.01  \\
2018-04-07 & 215.5 & $+$87.7  &  20.07 $\pm$ 0.05 & 18.87 $\pm$ 0.02 &  17.36 $\pm$ 0.01 &  16.58 $\pm$ 0.02 & 16.15 $\pm$ 0.02  \\
2018-04-12 & 220.5 & $+$92.7  &          --       & 18.91 $\pm$ 0.01 &  17.51 $\pm$ 0.01 &  16.67 $\pm$ 0.01 & 16.22 $\pm$ 0.01  \\
2018-04-15 & 223.5 & $+$95.7  &          --       &         --       &  17.53 $\pm$ 0.01 &  16.72 $\pm$ 0.01 & 16.30 $\pm$ 0.02  \\
2018-04-21 & 229.5 & $+$101.7 &          --       & 19.02 $\pm$ 0.01 &  17.60 $\pm$ 0.01 &  16.79 $\pm$ 0.01 & 16.37 $\pm$ 0.02  \\
2018-04-27 & 235.5 & $+$107.7 &          --       & 19.11 $\pm$ 0.03 &  17.70 $\pm$ 0.01 &  16.88 $\pm$ 0.01 & 16.44 $\pm$ 0.02  \\                      
2018-04-28 & 236.5 & $+$108.7 &          --       & 19.10 $\pm$ 0.03 &  17.68 $\pm$ 0.02 &  16.85 $\pm$ 0.03 & 16.43 $\pm$ 0.02  \\                           
2018-04-29 & 237.5 & $+$109.7 &          --       & 19.10 $\pm$ 0.02 &  17.68 $\pm$ 0.02 &  16.86 $\pm$ 0.01 & 16.46 $\pm$ 0.02  \\ 
2018-05-13 & 251.5 & $+$123.7 &          --       &        --        &  17.89 $\pm$ 0.01 &  17.02 $\pm$ 0.02 & 16.62 $\pm$ 0.03  \\
2018-05-19 & 257.5 & $+$129.7 &          --       &        --        &  17.98 $\pm$ 0.01 &  17.09 $\pm$ 0.01 & 16.68 $\pm$ 0.02  \\
2018-05-28 & 266.5 & $+$138.7 &          --       & 19.31 $\pm$ 0.03 &  18.13 $\pm$ 0.02 &  17.19 $\pm$ 0.02 & 16.79 $\pm$ 0.02  \\                      
2018-06-07 & 276.5 & $+$148.7 &  20.60 $\pm$ 0.05 & 19.47 $\pm$ 0.02 &  18.24 $\pm$ 0.01 &  17.30 $\pm$ 0.01 & 16.90 $\pm$ 0.02  \\                      
2018-06-11 & 280.5 & $+$152.7 &          --       &        --        &  18.32 $\pm$ 0.01 &  17.37 $\pm$ 0.01 & 16.91 $\pm$ 0.02  \\                        
2018-06-24 & 293.5 & $+$165.7 &          --       & 19.60 $\pm$ 0.02 &  18.49 $\pm$ 0.02 &  17.49 $\pm$ 0.01 & 17.05 $\pm$ 0.02  \\                            
2018-06-27 & 296.5 & $+$168.7 &          --       & 19.62 $\pm$ 0.02 &  18.41 $\pm$ 0.03 &  17.57 $\pm$ 0.01 & 17.12 $\pm$ 0.02  \\                           
2018-06-30 & 299.5 & $+$171.7 &          --       & --    $\pm$ --   &  18.59 $\pm$ 0.01 &  17.54 $\pm$ 0.01 & 17.11 $\pm$ 0.02  \\                           
2018-07-05 & 304.5 & $+$176.7 &          --       & 19.68 $\pm$ 0.03 &  18.59 $\pm$ 0.01 &  17.60 $\pm$ 0.01 & 17.20 $\pm$ 0.02  \\                           
2018-08-03 & 334.3 & $+$206.5 &          --       & 19.93 $\pm$ 0.04 &  19.02 $\pm$ 0.04 &  18.02 $\pm$ 0.02 & 17.60 $\pm$ 0.03  \\                                
2018-10-12 & 403.5 & $+$275.7 &          --       & --    $\pm$ --   &  19.85 $\pm$ 0.03 &  18.86 $\pm$ 0.04 & 18.60 $\pm$ 0.03  \\                         
2018-10-31 & 423.1 & $+$295.3 &          --       & 20.78 $\pm$ 0.03 &  20.17 $\pm$ 0.02 &  19.23 $\pm$ 0.05 & 19.01 $\pm$ 0.04  \\   
\hline                                                                                                                         
\end{tabular}\\
$^{\ast}$ With reference to the explosion date (JD~2458127.8).
\end{table*}

\begin{table*}[hbt!]
\centering
\caption{Swift/UVOT photometry for SN~2018gj}
\label{tab:Swiftphotometry}
\begin{tabular}{ccccccccc} 
\hline
Date        &  JD      & Phase$^{\ast}$    &  $UVW2$             & $UVM2$              & $UVW1$              & $UVU$        & $UVB$ & $UVV$     \\     
(yyyy-mm-dd)& 2458000+ &(d)                &  (mag)           & (mag)            & (mag)            & (mag)      & (mag)   & (mag)         \\  
\hline   
2018-01-16 &134.6 & $+$6.8 & 14.11 $\pm$ 0.03 & 13.92 $\pm$ 0.03 & 13.75 $\pm$ 0.03 & 13.70 $\pm$ 0.03 &14.82 $\pm$ 0.03& 14.72 $\pm$ 0.04 \\
2018-01-18&136.6 & $+$8.8 & 14.55 $\pm$ 0.03 & 14.33 $\pm$ 0.04 & 14.08 $\pm$ 0.03 &13.81 $\pm$ 0.03& 14.83 $\pm$ 0.03& 14.78 $\pm$ 0.04\\
2018-01-20&139.1 & $+$11.3 & 15.01 $\pm$ 0.04 & 14.98 $\pm$ 0.06 & 14.51 $\pm$ 0.04 &13.96 $\pm$ 0.04 &14.89 $\pm$ 0.04& 14.90 $\pm$ 0.06 \\
2018-01-22&141.4 & $+$13.6 & 15.50 $\pm$ 0.03 & 15.49 $\pm$ 0.04 & 14.91 $\pm$ 0.03 &14.20 $\pm$ 0.03&14.95 $\pm$ 0.03 &14.85 $\pm$ 0.04 \\
2018-01-24& 142.7 & $+$14.9 & -- & -- & -- &14.34 $\pm$ 0.03 &15.05 $\pm$ 0.03 &14.87 $\pm$ 0.05 \\
\hline                                                                                                                         
\end{tabular}\\
$^{\ast}$ With reference to the explosion date (JD~2458127.8).
\end{table*}

\begin{table*}[hbt!]
\centering
\caption{The $JHKs$ photometric magnitudes of SN~2018gj.}
\label{tab:KANATAphotometry}
\begin{tabular}{cccccc} 
\hline

Date        &  JD      & Phase$^{\ast}$    &  $Ks$             & $H$              & $J$            \\     
(yyyy-mm-dd)& 2458000+ &(d)                &  (mag)           & (mag)            & (mag)               \\ 
\hline 

2018-01-17	&	136.3	&	+8.5	&	14.05	$\pm$	0.09	&	--	&	14.28	$\pm$	0.03	\\
2018-01-19	&	138.4	&	+10.6	&	13.97	$\pm$	0.04	&	14.15	$\pm$	0.03	&	14.20	$\pm$	0.03	\\
2018-01-30	&	149.3	&	+21.5	&	13.88	$\pm$	0.03	&	14.08	$\pm$	0.03	&	14.20 $\pm$	0.02	\\
2018-02-03	&	153.3	&	+25.5	&	13.91	$\pm$	0.03	&	14.10	$\pm$	0.02	&	14.23	$\pm$	0.02	\\
2018-02-06	&	156.3	&	+28.5	&	13.85	$\pm$	0.03	&	14.08	$\pm$	0.02	&	14.13	$\pm$	0.02	\\
2018-02-07	&	157.3	&	+29.5	&	13.84	$\pm$	0.02	&	14.02	$\pm$	0.02	&	14.15	$\pm$	0.02	\\
2018-02-08	&	157.9	&	+30.1	&	13.83	$\pm$	0.03	&	14.05	$\pm$	0.02	&	14.16	$\pm$	0.02	\\
2018-02-14	&	164.3	&	+36.5	&	13.90	$\pm$	0.04	&	14.05	$\pm$	0.03	&	14.21	$\pm$	0.02	\\
2018-02-16	&	166.2	&	+38.4	&	14.00	$\pm$	0.04	&	14.08	$\pm$	0.02	&	14.23	$\pm$	0.02	\\
2018-02-21	&	171.2	&	+43.4	&	14.06	$\pm$	0.04	&	14.14	$\pm$	0.03	&	14.18	$\pm$	0.02	\\
2018-02-26	&	176.3	&	+48.5	&	13.92	$\pm$	0.04	&	14.09	$\pm$	0.03	&	14.24	$\pm$	0.02	\\
2018-02-27	&	177.0	&	+49.2	&	--	&	--	&	14.24	$\pm$	0.02	\\
2018-03-01	&	179.2	&	+51.4	&	14.01	$\pm$	0.03	&	14.19	$\pm$	0.02	&	14.32	$\pm$	0.02	\\
2018-03-09	&	187.3	&	+59.5	&	14.02	$\pm$	0.03	&	14.23	$\pm$	0.02	&	14.31	$\pm$	0.02	\\
2018-03-11	&	189.3	&	+61.5	&	13.96	$\pm$	0.04	&	14.12	$\pm$	0.03	&	14.29	$\pm$	0.02	\\
2018-03-12	&	190.3	&	+62.5	&	14.07	$\pm$	0.04	&	14.19	$\pm$	0.03	&	14.34	$\pm$	0.02	\\
2018-03-17	&	195.2	&	+67.4	&	14.15	$\pm$	0.04	&	14.30	$\pm$	0.02	&	14.41	$\pm$	0.02	\\
2018-03-22	&	200.1	&	+72.3	&	14.13	$\pm$	0.03	&	14.43	$\pm$	0.02	&	14.57	$\pm$	0.02	\\
2018-03-25	&	203.2	&	+75.4	&	14.30	$\pm$	0.04	&	14.51	$\pm$	0.03	&	14.66	$\pm$	0.02	\\
2018-03-28	&	206.1	&	+78.4	&	14.45	$\pm$	0.05	&	14.62	$\pm$	0.03	&	14.87	$\pm$	0.02	\\
2018-03-30	&	208.1	&	+80.4	&	14.72	$\pm$	0.04	&	14.85	$\pm$	0.03	&	15.06	$\pm$	0.02	\\
2018-04-01	&	210.2	&	+82.4	&	--	&	15.03	$\pm$	0.04	&	15.31	$\pm$	0.04	\\
2018-04-02	&	211.2	&	+83.4	&	14.98	$\pm$	0.08	&	15.09	$\pm$	0.04	&	15.38	$\pm$	0.04	\\
2018-04-07	&	216.3	&	+88.5	&	15.19	$\pm$	0.05	&	15.50	$\pm$	0.03	&	15.77	$\pm$	0.03	\\
2018-04-08	&	217.2	&	+89.4	&	--	&	15.47	$\pm$	0.04	&	15.74	$\pm$	0.03	\\
2018-04-12	&	221.3	&	+93.5	&	15.28	$\pm$	0.05	&	15.56	$\pm$	0.04	&	15.9	$\pm$	0.04	\\
2018-04-15	&	224.3	&	+96.5	&	15.25	$\pm$	0.09	&	15.62	$\pm$	0.05	&	15.93	$\pm$	0.04	\\
2018-04-17	&	226.3	&	+98.5	&	15.37	$\pm$	0.09	&	15.57	$\pm$	0.04	&	15.91	$\pm$	0.03	\\
2018-04-21	&	230.3	&	+102.5	&	--	&	15.69	$\pm$	0.05	&	15.94	$\pm$	0.04	\\
2018-04-22	&	231.3	&	+103.5	&	15.43	$\pm$	0.09	&	15.63	$\pm$	0.05	&	15.98	$\pm$	0.04	\\
2018-04-28	&	237.2	&	+109.4	&	15.48	$\pm$	0.10	&	15.79	$\pm$	0.07	&	16.08	$\pm$	0.06	\\
2018-04-30	&	239.2	&	+111.4	&	15.47	$\pm$	0.13	&	15.82	$\pm$	0.07	&	--	\\
2018-05-09	&	248.2	&	+120.4	&	15.74	$\pm$	0.11	&	16.08	$\pm$	0.06	&	16.36	$\pm$	0.03	\\
2018-05-11	&	250.1	&	+122.3	&	15.68	$\pm$	0.12	&	16.05	$\pm$	0.05	&	16.38	$\pm$	0.05	\\
2018-05-12	&	251.2	&	+123.4	&	--	&	--	&	16.42	$\pm$	0.05	\\
2018-05-21	&	260.1	&	+132.3	&	15.87	$\pm$	0.16	&	16.48	$\pm$	0.05	&	16.57	$\pm$	0.04	\\
2018-05-23	&	262.2	&	+134.4	&	15.91	$\pm$	0.12	&	16.50	$\pm$	0.07	&	16.66	$\pm$	0.04	\\
2018-05-31	&	270.2	&	+142.4	&	16.17	$\pm$	0.11	&	16.67	$\pm$	0.07	&	16.79	$\pm$	0.06	\\
2018-06-04	&	274.2	&	+146.4	&	16.23	$\pm$	0.27	&	--	&	16.76	$\pm$	0.08	\\
2018-06-16	&	286.1	&	+158.3	&	16.84	$\pm$	0.20	&	17.02	$\pm$	0.12	&	16.91	$\pm$	0.04	\\
2018-07-12	&	312.1	&	+184.3	&	--	&	17.64	$\pm$	0.23	&	--	\\
2018-08-01	&	332.0	&	+204.2	&	--	&	--	&	17.97	$\pm$	0.09	\\
\hline
\end{tabular}\\
$^{\ast}$ With reference to the explosion date (JD~2458127.8).
\end{table*}

\begin{table}[hbt!]
\centering
\caption{Log of spectroscopic observations of SN~2018gj.} 
\begin{tabular}{cccc} \hline
   Date          & JD         &Phase    & Range       \\        
(yyyy-mm-dd)     & (2458000+) &(d)      & (\AA)       \\        
\hline
2018-01-14        & 132.5   & $+$4.7      &  3500--9250 \\     
2018-01-16        & 134.5   & $+$6.7      &  3500--9250 \\     
2018-01-18        & 136.5   & $+$8.7      &  3500--9250 \\     
2018-01-21        & 139.5   & $+$11.7     &  3500--9250 \\     
2018-01-24        & 142.5   & $+$13.7     &  3500--9250 \\     
2018-01-25        & 143.5   & $+$15.7     &  3500--9250 \\     
2018-01-27        & 145.5   & $+$17.7     &  3500--9250 \\     
2018-02-02        & 151.5   & $+$23.7     &  3500--9250 \\     
2018-02-03        & 152.5   & $+$24.7     &  3500--9250 \\     
2018-02-06        & 155.5   & $+$27.7     &  3500--9250 \\     
2018-02-13        & 162.5   & $+$34.7     &  3500--9250 \\       
2018-02-16        & 165.5   & $+$37.7     &  3500--9250 \\     
2018-02-18        & 167.5   & $+$39.7     &  3500--9250 \\     
2018-02-20         & 169.5   & $+$41.7     &  3500--9250 \\     
2018-02-25        & 174.5   & $+$46.7     &  3500--9250 \\      
2018-03-14        & 191.5   & $+$63.7     &  3500--9250 \\             
2018-03-15        & 192.5   & $+$64.7     &  3500--9250 \\     
2018-03-18        & 195.5   & $+$67.7     &  3500--9250 \\     
2018-03-23        & 200.5   & $+$72.7     &  3500--9250 \\     
2018-03-30        & 207.5   & $+$79.7     &  3500--9250 \\     
2018-04-02        & 210.5   & $+$82.7     &  3500--9250 \\     
2018-04-07        & 215.5   & $+$87.7     &  3500--9250 \\       
2018-04-15        & 223.5   & $+$95.7     &  3500--7800 \\       
2018-04-21        & 229.5   & $+$101.7    &  3500--9250 \\     
2018-04-27        & 235.5   & $+$107.7    &  3500--9250 \\     
2018-05-01         & 239.5   & $+$111.7    &  3500--9250 \\             
2018-05-28        & 266.5   & $+$138.7    &  3500--9250 \\     
2018-06-11        & 280.5   & $+$152.7    &  3500--7800 \\      
2018-06-29      & 297.5   & $+$168.7    &  3500--9250 \\  
 2018-10-04        & 396.1   & $+$268.3    &  5200--9250 \\      
 2018-10-31       & 423.1   & $+$295.3    &  3500--7800 \\      
\hline
\end{tabular} \\
$^{\ast}$ With reference to the explosion date (JD~2458127.8). 
\label{tab:HCTspec}
\end{table}

\begin{table*}[hbt!]
\centering
\caption{Type II SN data used in comparison and estimating mean colour evolution.}
\label{tab:SampleColor}
\begin{tabular}{llcllcllc} 
\hline
S.No        &  SN      & Reference   &  S.No.        & SN          & Reference             & S.No.       & SN       & Reference \\
\hline
1 &1992H &\cite{1996AJ....111.1286C} & 16 &2006au &\cite{2012AA...537A.140T} & 31 &2013by &$\dagger$\\
2& 1992af&* &17 & 2007it &* & 32 & 2013fs&$\dagger$ \\
3& 1992ba &*&18 & 2007pk& \cite{2013AA...555A.142I}&33 & LSQ13dpa&$\dagger$ \\
4& 1997D &\cite{2003ApJ...582..905H} &19 & 2008gz& \cite{2011MNRAS.414..167R} &34 & 2013hj&\cite{2016MNRAS.455.2712B} \\
5& 1999em & *&20 &2008in &* &35 &2014G &$\dagger$\\
6& 1999gi& \cite{2002AJ....124.2490L}&21 & 2009E&\cite{2012AA...537A.141P} &36 & 2014cx& $\dagger$\\
7& 2000cb&\cite{2011MNRAS.415..372K} &22 & 2009N&* &37 & 2014dw& $\dagger$\\
8& 2002hx& *&23 & 2009bw& \cite{2012MNRAS.422.1122I} &38 & ASASSN-14ha & $\dagger$\\
9& 2003gd& *&24 & 2009ib &\cite{2015MNRAS.450.3137T} &39 & 2016X&\cite{2018MNRAS.475.3959H} \\
10&2004dj & \cite{2006AJ....131.2245Z}&25 & 2009md& \cite{2011MNRAS.417.1417F} &40 & 2016bkv& \cite{2018ApJ...859...78N} \\
11& 2004et&\cite{Sahu2006_2004et} &26 & 2012A& \cite{2012MNRAS.422.1122I} &41 & 2017eaw&\cite{2018AstL...44..315T} \\
12& 2004fx& *&27 & 2012aw& \cite{2013MNRAS.433.1871B}&42 & 2018ivc&\cite{2020ApJ...895...31B} \\
13& 2005af& *&28 & 2012ec&\cite{2015MNRAS.448.2312B} &43 & 2018zd& \cite{2020MNRAS.498...84Z} \\
14&2005cs &\cite{Pastorello2006_2005cs} &29 &2013K & \cite{2018MNRAS.475.1937T} &44 & 2020jfo& \cite{2022ApJ...930...34T} \\ 
15& 2006V& \cite{2012AA...537A.140T} &30 & 2013ab&$\dagger$ & - &- &- \\

\hline  
\hline                                                                                                                         
\end{tabular}\\
* \cite{2014Anderson}, $\dagger$\cite{2016MNRAS.459.3939Valenti}
\end{table*}

\newpage 

\section{EPM distance to SN~2018gj}
\label{appendix:EPM}
\begin{figure}[hbt!]
	 \resizebox{\hsize}{!}{\includegraphics{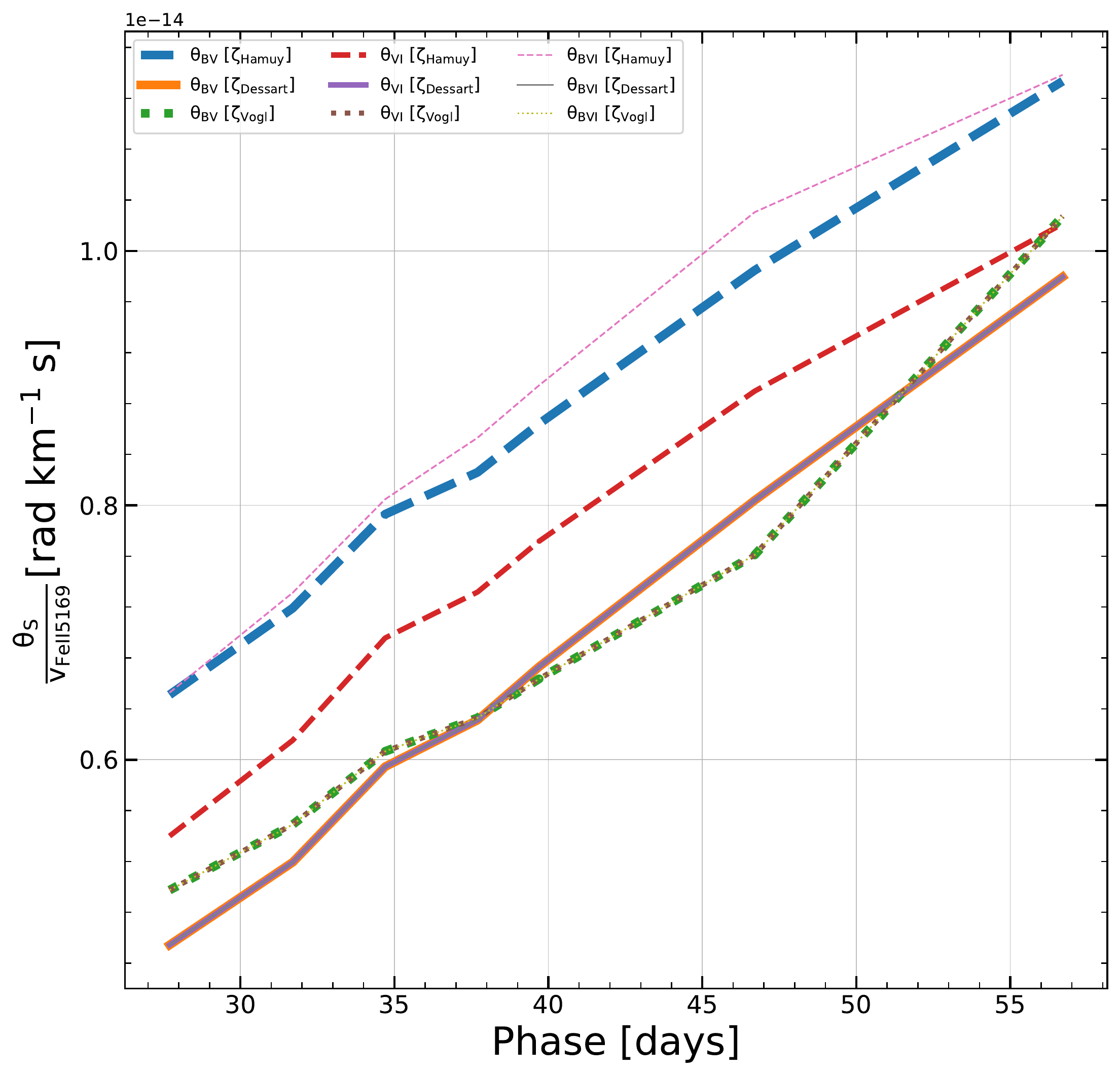}}
    \caption{EPM distance estimates for various filter sets and different dilution factors from \citet{hamuy2001_1999em}, \citet{DH2005_EPMdilution} and \citet{Vogl2019_EPMDilution}}
    \label{fig:EPM_values}
\end{figure}

The implementation of expanding photosphere method (EPM) was followed as per the details given in \citep{hamuy2001_1999em} and \citep{DH2005_EPMdilution}. This formalism involves the measurements of two radii associated with SN i) a photometric angular radius ($\theta$)  and ii) a spectroscopic physical
radius (R). With the aid of these two radii, the distance to SN could be derived. The angular radius ($\theta$) is given as:
$$\rm \theta = \frac{R}{D} = \sqrt{\frac{f_\lambda}{\pi B_\lambda (T) 10^{-0.4A(\lambda)}\zeta_\lambda^2 }}, $$ where, D is the distance to the SN. $B_\lambda(T)$ is the Planck function at the colour temperature of the blackbody radiation, $f_\lambda$ is the apparent flux density, A($\lambda$) is the dust extinction, and $\zeta_\lambda$ is the dilution factor to account for the deviation from a black body \citep{hamuy2001_1999em}. The above equation could be transformed in terms of apparent magnitudes ($m_\lambda$) for multiband photometry as:
$$\rm m_\lambda= -5 log(\zeta_\lambda) - 5 log(\theta) + A_\lambda + b_\lambda(T) ,$$
Now, for different filters set (S) above equation was minimized with $b_\lambda$ values being taken from \citet{hamuy2001_1999em}, and dilution factors have been considered from three different works viz. \citet{hamuy2001_1999em}, \citet{DH2005_EPMdilution} and \citet{Vogl2019_EPMDilution}. The quantity that was minimized is as follows:
$$\rm \varepsilon = \sum_{\lambda \epsilon S} \left[ m_\lambda  + log(\theta_S \zeta_S) - A_\lambda - b_\lambda(T_S) \right]^2 $$

Finally, using the expansion velocity ($v$) measured using spectra could be used in the following equation:
 $$ \frac{\theta_i}{v_i} \approx \frac{(t_i - t_0)}{D}$$, where subscript i implies for each epoch available. A straight line could be fit for multiple epochs, and the resulting slope is used to get estimates of the distance.

\newpage
\section{Scaling relation for probable progenitor}
\label{appendix:scaling}
Scaling relations obtained in the \citet{2019ApJ...879....3GGOLD} that give a set of probable explosions which could yield observed bolometric light curve can be used for initial model guess in MESA models. These relations are solved to obtain $\rm M_{ej}$ and $\rm E_{exp}$ as a function of $\rm M_{Ni}$, $\rm L_{50}$, $\rm t_p$, and $\rm R$ as:
\begin{equation}
  \rm   \log(E_{51}) = -0.728 + 2.148\log(L_{42}) -0.280\log(M_{Ni})+2.091\log(t_{p,2})-1.632\log(R_{500})
\end{equation}, and

\begin{equation}
   \rm \log(M_{10})=-0.947 + 1.474\log(L_{42})-0.518\log(M_{Ni})+3.867\log(t_{p,2})-1.120\log(R_{500})
\end{equation}
where $\rm E_{51}$ is explosion energy in the units $\rm 10^{51}~ergs$, $\rm M_{10}$ is the ejecta mass in the units $\rm 10~M_\odot$, $\rm M_{Ni}$ has $\rm M_\odot$ unit, $\rm L_{42}=$ Luminosity at 50 days/$\rm 10^{42}~erg~s^{-1}$, $\rm R_{500}=$ Progenitor Radius/500~$\rm R_\odot$, and $\rm t_{p,2}=$ Plateau length/100~d.

\begin{figure}[hbt!]
	 \resizebox{\hsize}{!}{\includegraphics{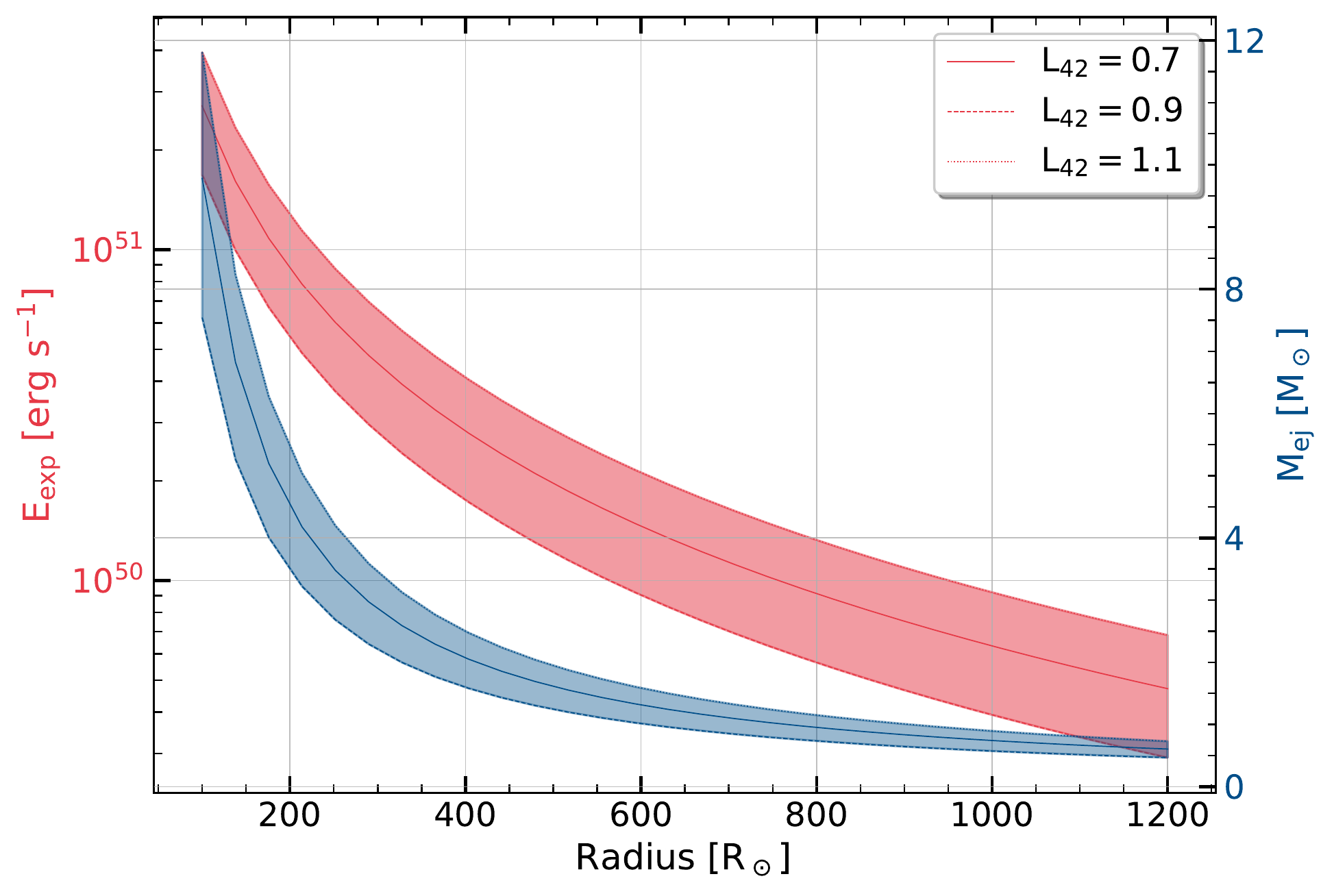}}
    \caption{Scaling relations \citep{2019ApJ...879....3GGOLD} in the context of Type IIP SNe as applicable to SN~2018gj.}
    \label{fig:Scaling_Relations}
\end{figure}

\section{Model Light curves and velocity evolution for 19 \texorpdfstring{$\rm M_\odot$}~~Models}
\label{appendix:19M}

\begin{figure}[h!]
	 \resizebox{\hsize}{!}{\includegraphics{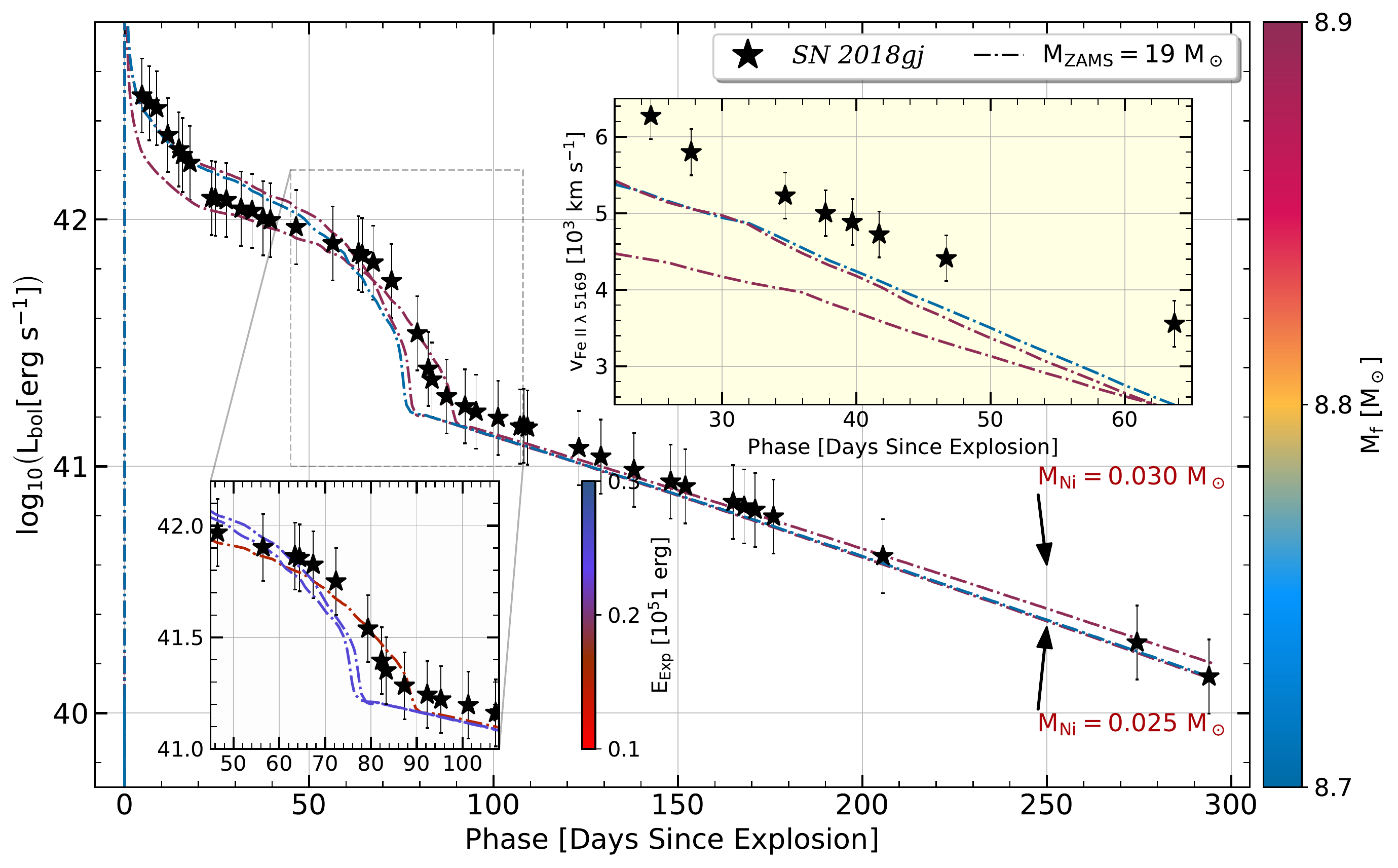}}
    \caption{Variations in 19 $\rm M_\odot$ ZAMS model using different parameters to achieve a shorter plateau length. Zoomed out a plot in the bottom left shows the variation in explosion energy for different model light curves around the plateau transition. The second plot in the right inset shows the corresponding Fe 5169 velocities obtained using models. All the 19 $\rm M_\odot$ models underestimate the velocity evolution.}
    \label{fig:mesa_stella19m}
\end{figure}


\bibliography{SN2018gj}{}
\bibliographystyle{aasjournal}



\end{document}